\documentclass[prl,twocolumn,preprintnumbers,amsmath,amssymb,superscriptaddress]{revtex4}

\usepackage{enumerate, amsmath}
\usepackage{epsf,graphicx,color}
\usepackage{amssymb}
\usepackage{epstopdf}
\usepackage{subfigure}
\usepackage{color}
\usepackage{setspace}

\begin{document}

\bibliographystyle{prsty}

%\pagenumbering{arabic}

%%%%%%   TITLE AND AUTHORS   %%%%%%

\title{
Temperature Dependence of Nonlinear Susceptibilities in an
Infinite Range Interaction Model
}
\author{Pradeep Kumar and Christopher E.\ Wagner}
\affiliation{Department of Physics, University of Florida 32611-8440, USA}

%%%%%%%%%%   ABSTRACT   %%%%%%%%%%%

\begin{abstract}
We present a model to probe metamagnetic properties in systems with an
arbitrary number of interacting spins.  Thermodynamic properties such as
the magnetization per particle $m(B,T,N)$, linear susceptibility
$\chi_1(T)$, and nonlinear susceptibilities $\chi_3(T)$ and $\chi_5(T)$
were calculated.  The model produces a different magnetic response for
$N$ particles when comparing to $N - 1$ particles for small $N \sim 1$.
For an even number of particles, the susceptibilities show maxima in
their temperature dependence.  An odd number produces an additional free
spin response that dominates at low temperatures.  This free spin
response also produces a step in the magnetization per particle at
$B = 0$ for odd $N$.  The magnetization shows $N/2$ steps at
$\gamma B_c/J = n$ with integer $n$ for even $N$ and $(N-1)/2$
additional steps at with integer $n$ for odd $N$.  Small clusters
respond with metamagnetism in an otherwise isotropic spin space, while
the large clusters show no metamagnetism.
\end{abstract}

%%%%%%%%%%   BODY   %%%%%%%%%%%

%following command places the title, authors, and abstract.
\maketitle

\section{1. Introduction}

Metamagnetism \cite{Stryjewski:77,Levitin:88,Goto:01} is identified as when
the magnetization rapidly rises at a critical magnetic field.  Typically this
happens at low temperatures and the critical field is nearly insensitive to
the temperature.  The transition broadens with increased temperature.  In a
transition metal antiferromagnet, the transition is described
\cite{Stryjewski:77} as a first order spin flop transition.  When the
magnetic field is applied along some unpreferred direction, the low field
response is weak because the spins are locked along the preferred direction.
The spins line up along the magnetic field when the field exceeds the
anisotropy energy.

It was proposed \cite{wohlfarth:62,Aoki:13} that for metals a first order
phase transition occurs.  When considering an expansion of the Ginzburg Landau
free energy in terms of the magnetization, this transition is caused by a
negative fourth order term due to the density of states of the material.
The metal would then have a metamagnetic phase transition similar to a
liquid-gas phase transition.

Many materials \cite{Rost:09, Thomas:16} show a sharp metamagnetism only at
$T = 0$, contrary to the Wohlfarth-Rhodes narrative.  With a $T > 0$, the
transition is no longer discontinuous and smoothly disappears.  This is
consistent with a quantum phase transition \cite{Millis:02,DHoker:10},
thermodynamically analogous to the ferromagnetic transition.  Where the
ferromagnetic transition occurs at $B = 0$ and $T = T_c$, the metamagnetic
transition occurs at $B = B_c$ and $T = 0$.

Metamagnetism in an insulating antiferromagnet is easily discussed in terms
of an anisotropic exchange Hamiltonian.  The anisotropy derives from
crystalline electric fields and depends on the field orientation with
respect to the lattice structure.  In a metal such as the heavy fermions,
the sensitivity to the lattice structure is less clear in experiments.
However, a discrete level structure whose crossing represents the
metamagnetism should be describable in terms of a spin Hamiltonian.  We
therefore pick an example (see another example in Ref.\ \cite{Bak:82}) and
study its consequences.

The recent study \cite{Shivaram:14} by Shivaram et al., starting with the
measurement of metamagnetism for heavy fermion compounds, notes certain
correlations between the observables.  It was already noted by Goto et al.\
\cite{Goto:01} that the linear susceptibility showed a peak at a temperature
$T_1$ and that the critical field $B_c$ scaled with the inverse susceptibility
at that maximum.  Hirose et al.\ \cite{Hirose:11} pointed out that critical
field followed the susceptibility peak temperature.  Shivaram et al.\ also
noted the correlation in the peak temperatures of the nonlinear
susceptibilities.  They seemed to favor a single energy scale in these
phenomena.  Thermodynamics connects the temperature dependence of
magnetization to the field dependence of the specific heat or the sound
velocity.  Here, we find similar results for an otherwise general Hamiltonian.
The series of compounds CeMIn$_5$ show a Curie law susceptibility at low
temperatures as reported by Thamizhavel et al.\ \cite{Thamizhavel:15}.
B.\ Shivaram \cite{Shivaram:15} has noted several instances of metamagnetism
in different contexts.  For a review of molecular magnets see also
Ref.\ \cite{Schnack:13}.

This paper presents an overview of a model describing a cluster of $N$
fermionic (spin $s = 1/2$) spins interacting through an infinite range antiferromagnetic
interaction $J$.  These interacting spins are influenced by an external
magnetic field $B$ in the direction of the spin quantization axis $z$.  The
model is meant to solely study the spin contribution to metamagnetism.
Using the dimensionless variables $\tau = k_B T/J$ and $b = \gamma B/J$, the
principal results are summarized here:

\begin{enumerate}

\item The responses for even and odd number of particles are qualitatively
different.  There are $(N + n_0)/2$ steps in the magnetization with
$n_0 = N \mod 2$ describing the ``oddness'' of $N$ (for even $N$, $n_0 = 0$
and for odd $N$, $n_0 = 1$).  The steps occur at critical field values
\[
B_c = \left( \frac{2n + n_0}{2} \right) \frac{J}{\gamma} \quad \quad n = 1,2,...,\frac{N-n_0}{2}
\]
The magnetization per particle changes by $1/N$ at each $B_c$.  When $N$ is
odd, there is an additional step at $B_c = 0$ due to a free spin response
which changes the magnetization per particle by $1/2N$.  The fully saturated
magnetization per particle is $m = 1/2$ for all $N$.

\item For odd $N$, the ground state is a Kramer's doublet leading to a free
spin response (see discussion after Eq.\ 3.1 in Ref.\ \cite{Alcaraz:88}); a Curie law contribution to the total susceptibility.

\item The nonlinear susceptibilities are defined as
\[
M = \chi_1(T) B + \chi_3(T) B^3 + \chi_5(T) B^5.
\]
The third and fifth order susceptibilities are negative at high temperatures.
The susceptibilities rise at low temperatures and show a maximum at a
characteristic temperature.  For even $N$, they all go to zero at $T = 0$.
For odd $N$, there is a free spin response which dominates at low temperatures.
Shivaram et al.\ \cite{Shivaram:14} studied another basic Hamiltonian
$H = \Delta S_z^2 - \gamma B S_z$ showing similar results.

\item The specific heat as a function of temperature at $b = 0$ is
Schottky-like.  It rises exponentially at low temperatures and decays as
$\tau^{-2}$ at high temperatures.  As a function of magnetic field at low
temperatures, the specific heat has a minimum at the critical fields
buttressed by peaks both below and above the critical field.  As the
temperature increases, the minima stay fixed but the peaks move out.

\end{enumerate}

The rest of the paper is organized as follows: Sec.\ 2 focuses on the details
of the infinite range exchange interaction model.  The partition function and
the observables are discussed for a generalized particle number $N$, commenting
on the difference in the even $N$ and odd $N$ cases. Sec.\ 3 presents the
results in the small $N$ limit ($N = 2,3$), and Sec.\ 4 discusses the large
$N$ limit ($N = 23, 24$) and the thermodynamic limit ($N \rightarrow \infty$).
Sec.\ 5 contains a summary of the results and a discussion of the limitations
of the model. 

\section{2. Model}

In the model considered here, each spin interacts with all other spins.  The
Hamiltonian is
\begin{equation}
H = J \sum_{i < j} \mathbf s_i \cdot \mathbf s_j - \gamma \mathbf B \cdot \mathbf S.
\label{eqn:H}
\end{equation}
Here $J$ is an antiferromangeitc exchange interaction of infinite range,
$\gamma$ is the gyromagnetic ratio, $B$ is the external field, and
$\mathbf S = \sum_i \mathbf s_i$ is the total spin of the system.  The
eigenenergies of the system are calculated using the identity:
\begin{equation}
2 \sum_{i<j} \mathbf s_i \cdot \mathbf s_j = \left( \sum_i \mathbf s_i \right)^2 - \sum_i s_i^2 = S^2 - \frac{3}{4} N.
\label{eqn:identity}
\end{equation}
This produces the eigenenergies
\begin{equation}
\lambda = \frac{J}{2} S(S + 1) - \frac{3N}{8} J - \gamma B \mu,
\label{eqn:einrgy}
\end{equation}
where $\mu$ is the $z$ component of spin and $S$ is the quantized spin
eigenvalue.  The range of values for $\mu$ and $S$ depend on whether there are
an even or odd number of particles.  For generalization, it is useful to define
the variable $n_0 = N \mod 2$, same as above.  The range of values for $\mu$
and $S$ can then be written $-S \le \mu \le S$ and
$\frac{1}{2} n_0 \le S \le N/2$, with all values spaced by an integer.  The
partition function can then be written from these eigenenergies
\begin{equation}
Z = e^{3JN/8} \sum_{S = \frac{1}{2} n_0}^{N/2} e^{-S(S + 1)/2\tau} \sum_{\mu = -S}^S e^{\mu b/\tau}.
\label{eqn:Z}
\end{equation}
Here $\tau = k_B T/J$ and $b = \gamma B/J$ are dimensionless variables
describing the temperature and field.  The prefactor $\exp(3JN/8)$ is a
constant multiple that will cancel when calculating thermal properties, and
will thus be dropped from here on.  Finally the partition function leads to
the Gibbs' free energy such that $F = -k_B T \ln Z$, from which comes the
thermodynamic properties: magnetization $M = - \frac{\partial F}{\partial B}$,
specific heat $C = - T \frac{\partial^2 F}{\partial T^2}$ and pressure
$P = - \frac{\partial F}{\partial V}$.  For the pressure there is no explicit volume dependence.  The results in this paper assume a volume dependence on the coupling strength $J(V) = g V$ leading to the pressure $P = - \left( \frac{\partial J}{\partial V} \right) \frac{\partial F}{\partial J} = -g \frac{\partial F}{\partial J}$.  For simplicity all discussion of pressure is in units of $g$.

To more easily evaluate the thermodynamic properties, the partition function
can be rewritten by changing the order of the summations.  To simplify the
expression, the field-independent function
$A_\mu(N) = \sum_{S = \mu}^{N/2} e^{-S(S + 1)/2\tau}$ is used.  The even and
odd partition functions are then,
\begin{align}
Z(N = 2n) &= A_0(N) + 2 \sum_{\mu = 1}^{N/2} A_\mu(N) \cosh \left( \frac{\mu b}{\tau} \right), \nonumber \\
Z(N = 2n + 1) &= 2 \sum_{\mu = 1/2}^{N/2} A_\mu(N) \cosh \left( \frac{\mu b}{\tau} \right).
\label{eqn:modZ}
\end{align}
From here the magnetization $M$ for each becomes,
\begin{align}
\frac{M(N = 2n)}{\gamma} &= \frac{2 \sum_{\mu = 1}^{N/2} \mu A_\mu(N) \sinh(\mu b/\tau)}{A_0(N) + 2 \sum_{\mu = 1}^{N/2} A_\mu(N) \cosh(\mu b/\tau)}, \nonumber \\
\frac{M(N = 2n + 1)}{\gamma} &=  \frac{\sum_{\mu = 1/2}^{N/2} \mu A_\mu(N) \sinh(\mu b/\tau)}{\sum_{\mu = 1/2}^{N/2} A_\mu(N) \cosh(\mu b/\tau)}.
\label{eqn:M}
\end{align}
The nonlinear susceptibilities follow from these expressions.  Likewise the
other thermodynamic observables such as specific heat or pressure (the field
dependent part, using the implicit volume dependence of $J$) can be obtained
using the other thermodynamic derivatives.
\begin{figure*}
\includegraphics[width=0.32\linewidth]{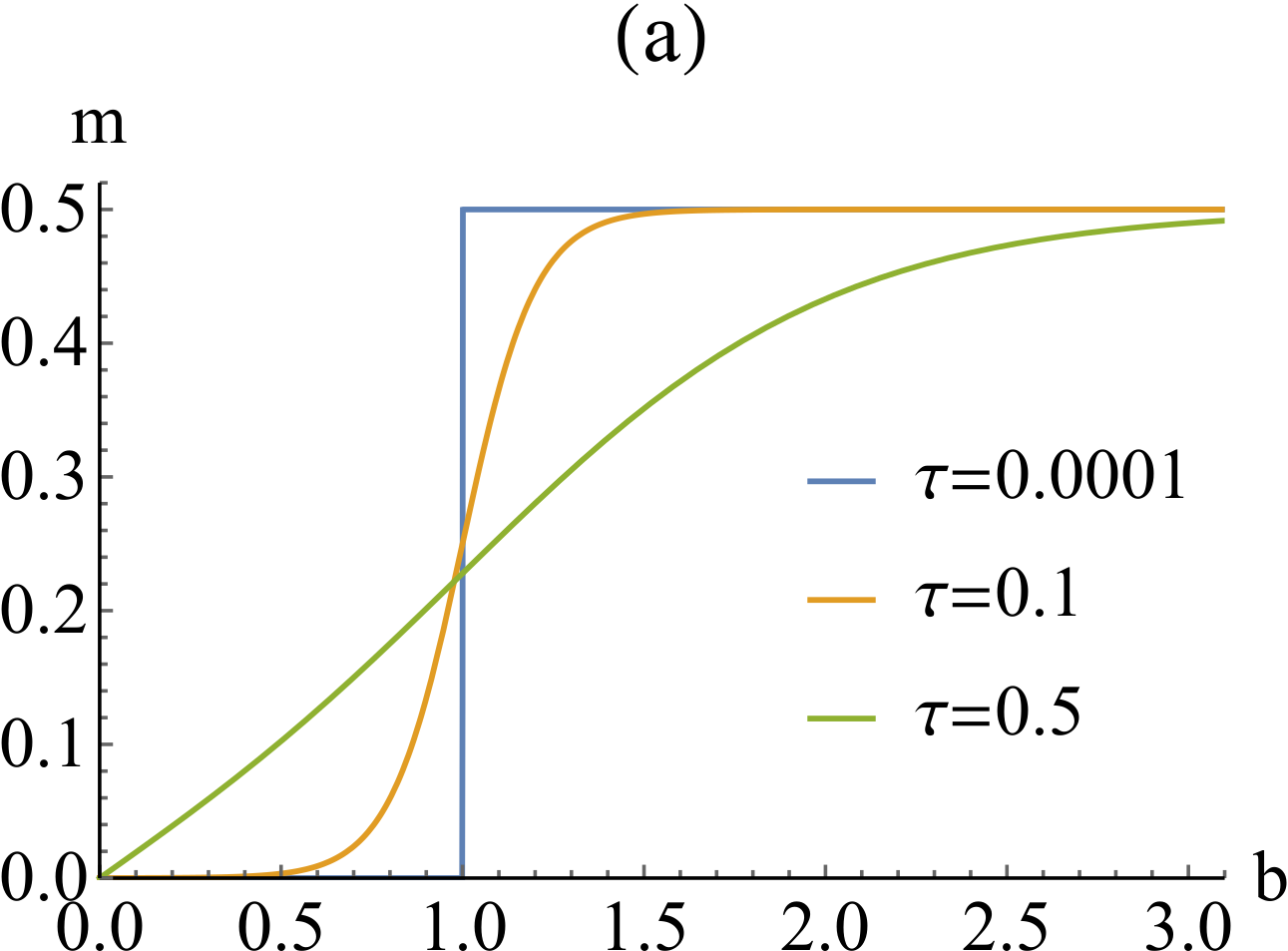}
\includegraphics[width=0.32\linewidth]{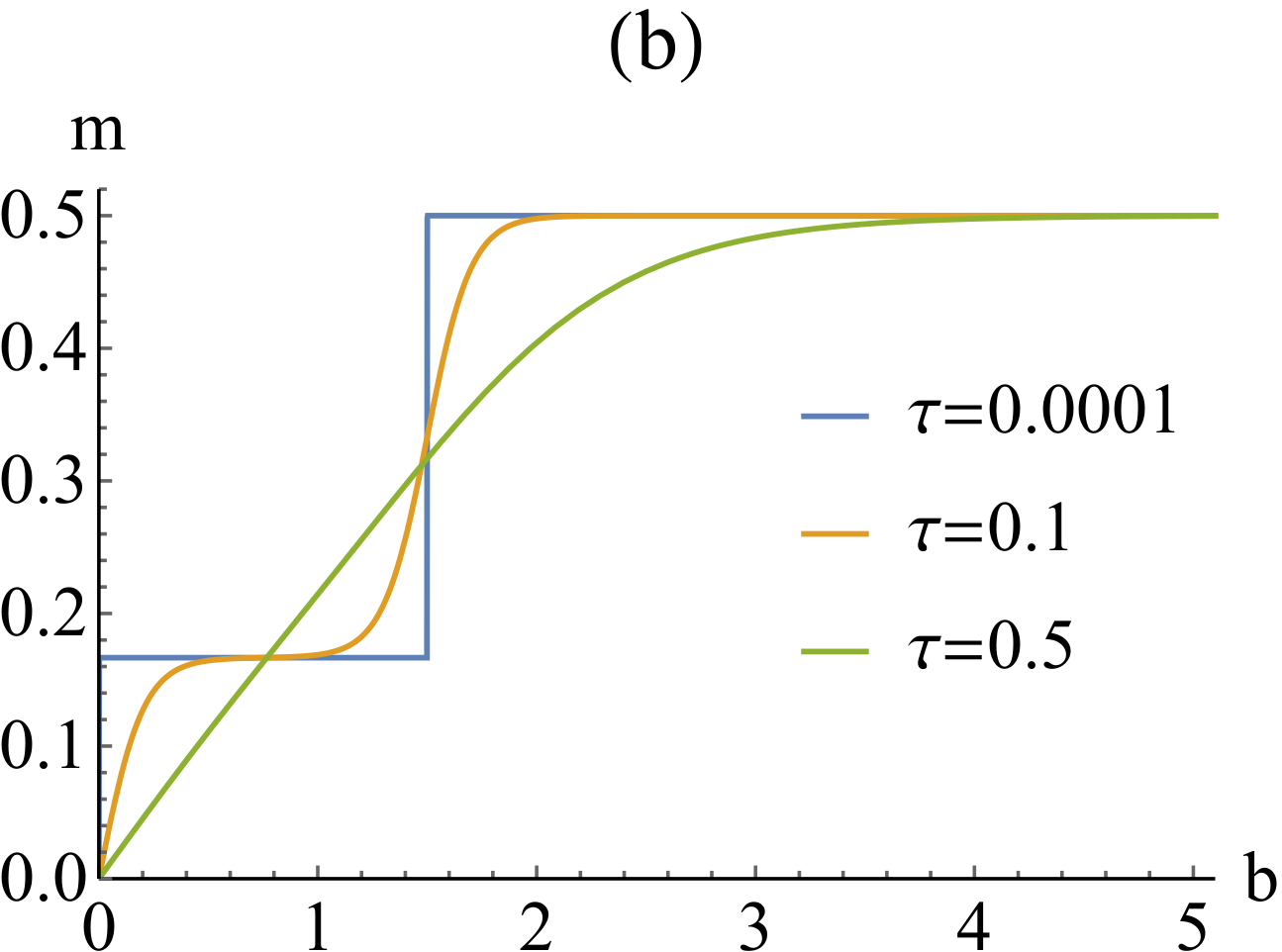}
\includegraphics[width=0.32\linewidth]{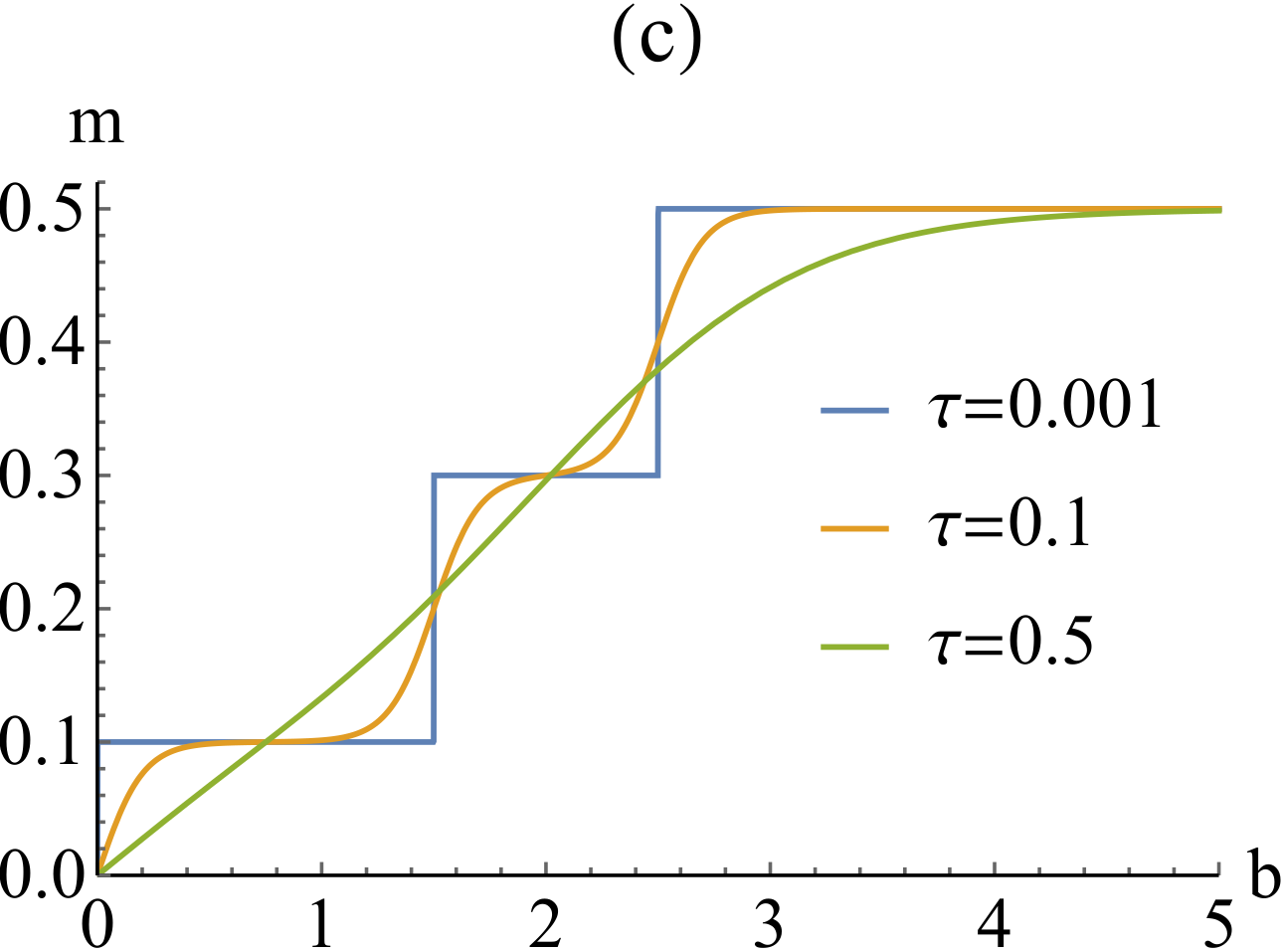}
\includegraphics[width=0.32\linewidth]{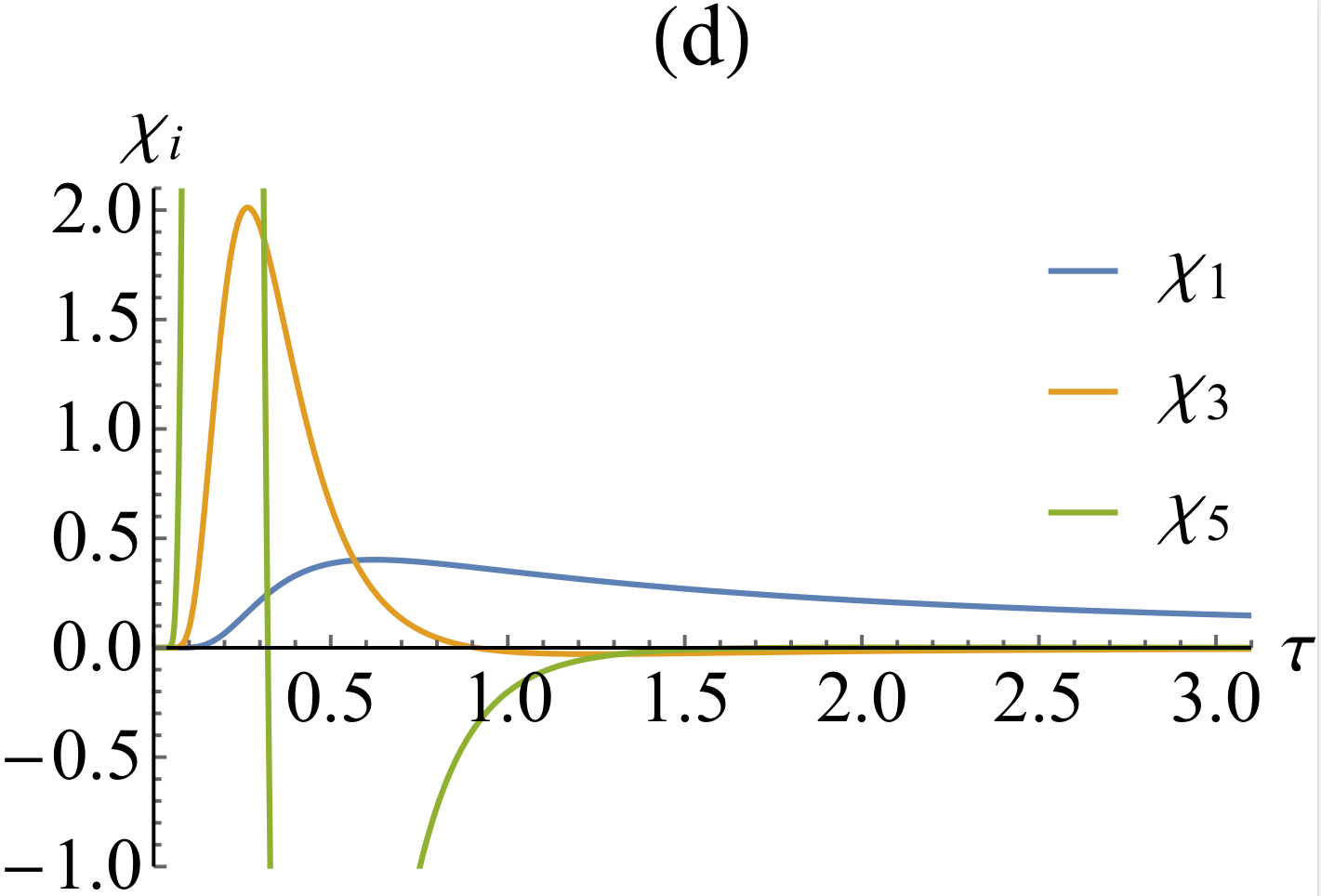}
\includegraphics[width=0.32\linewidth]{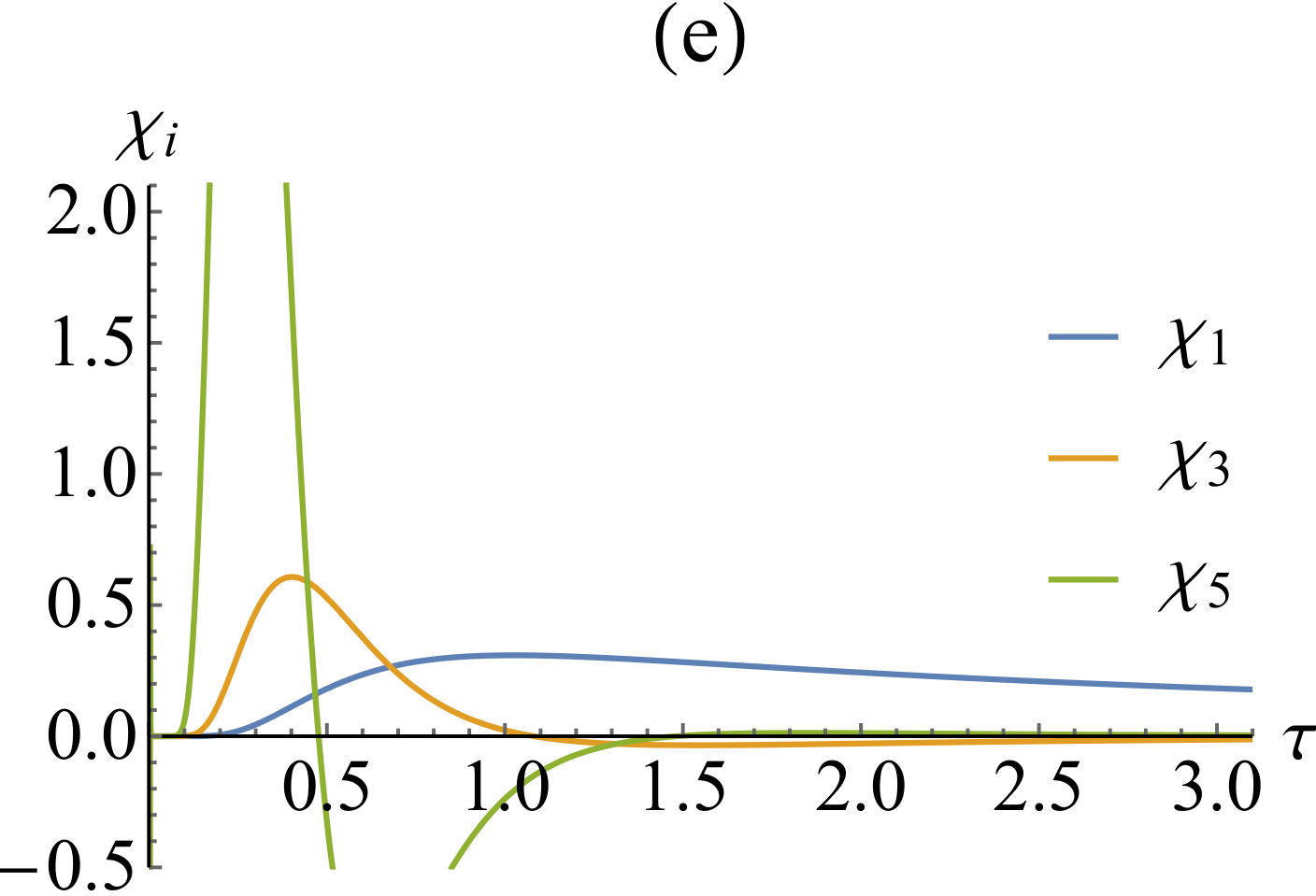}
\includegraphics[width=0.32\linewidth]{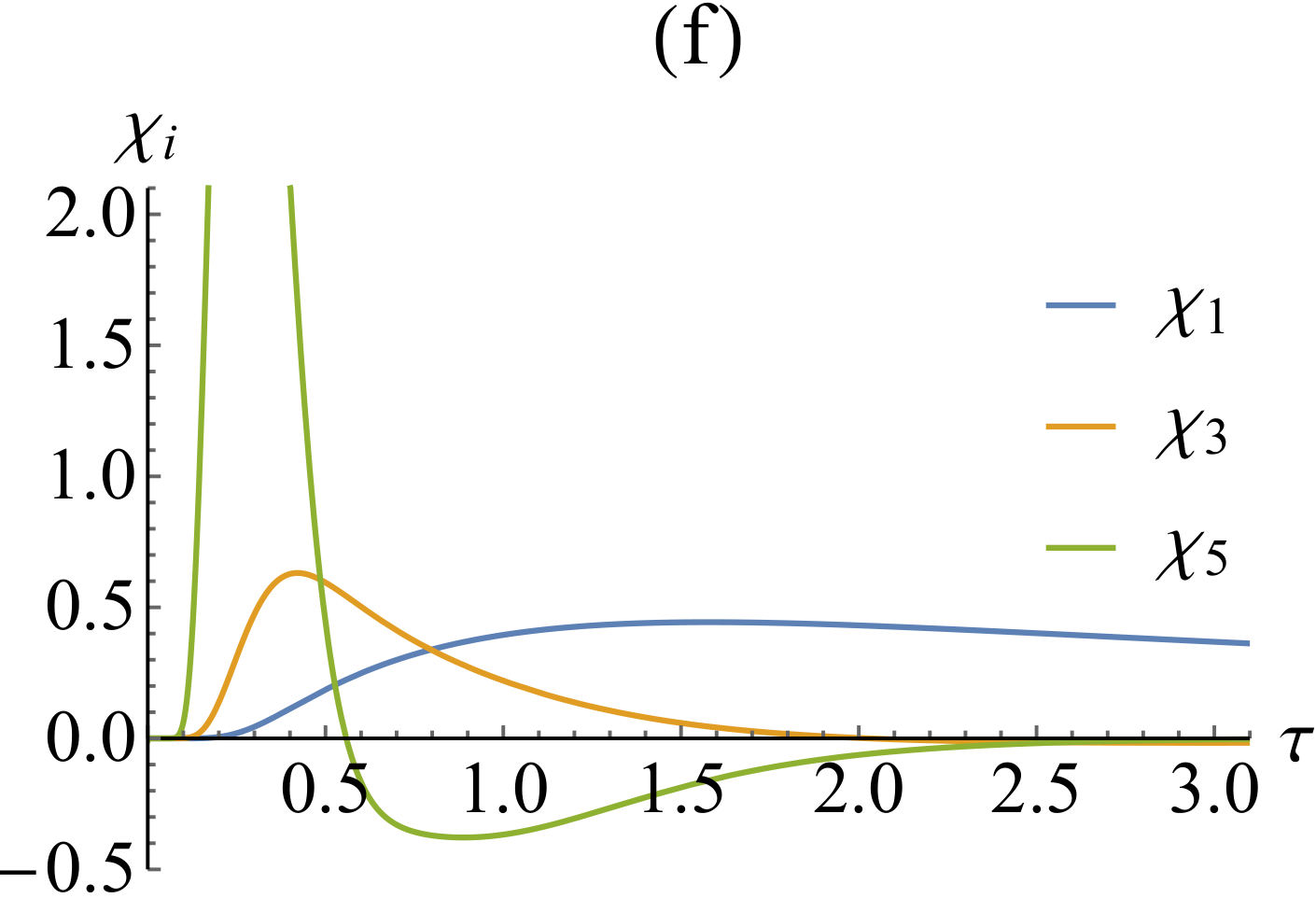}
\includegraphics[width=0.32\linewidth]{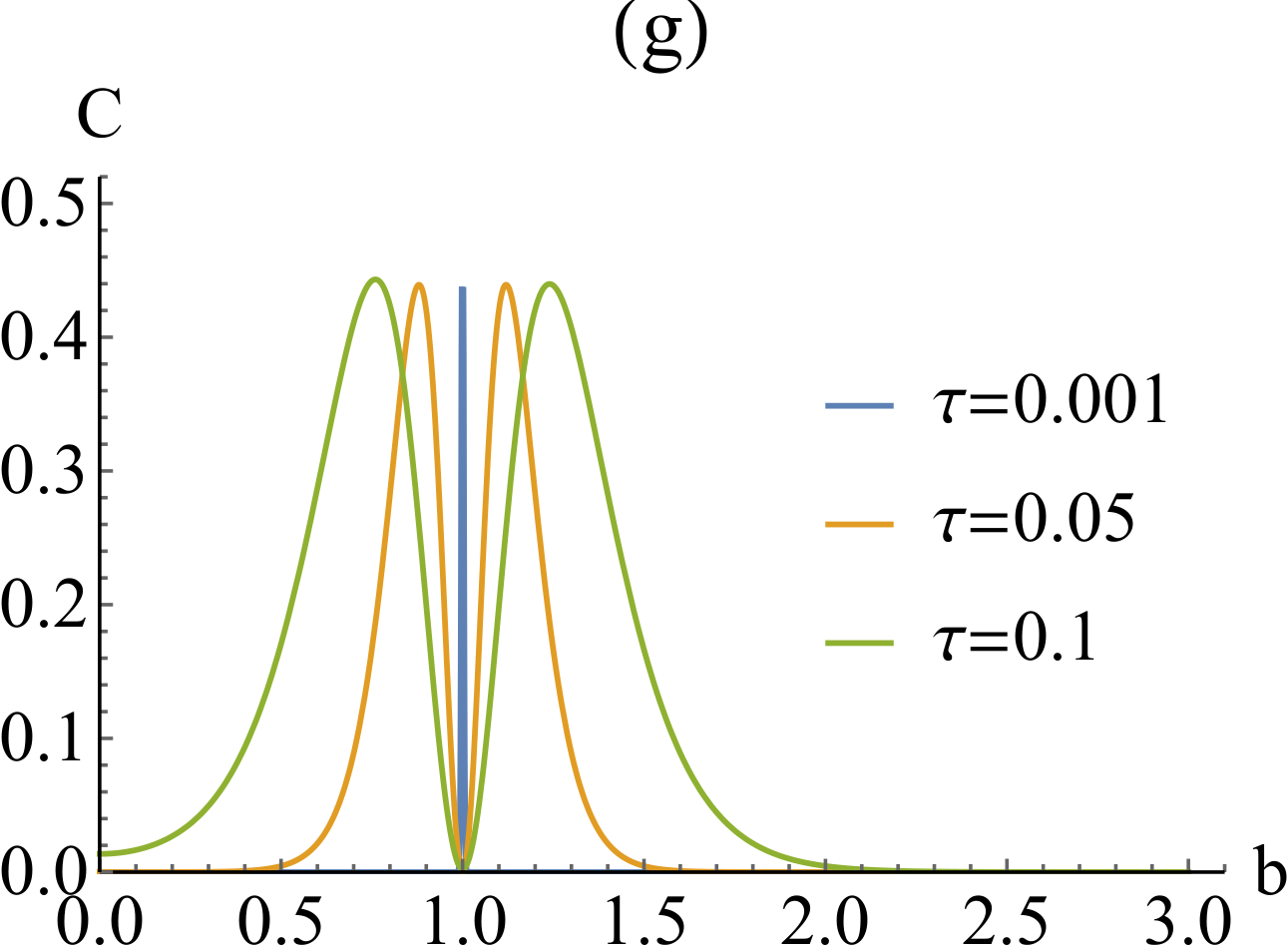}
\includegraphics[width=0.32\linewidth]{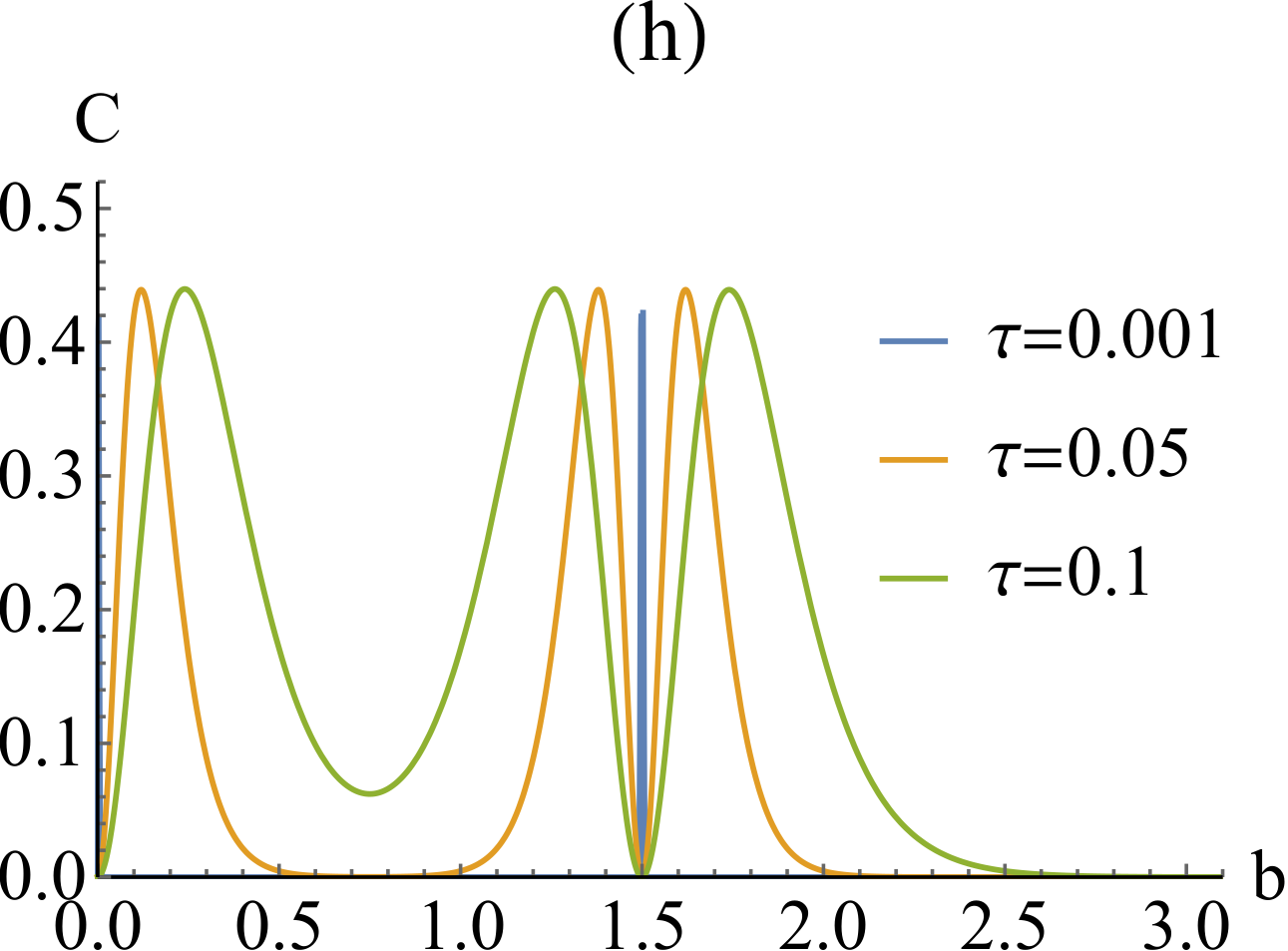}
\includegraphics[width=0.32\linewidth]{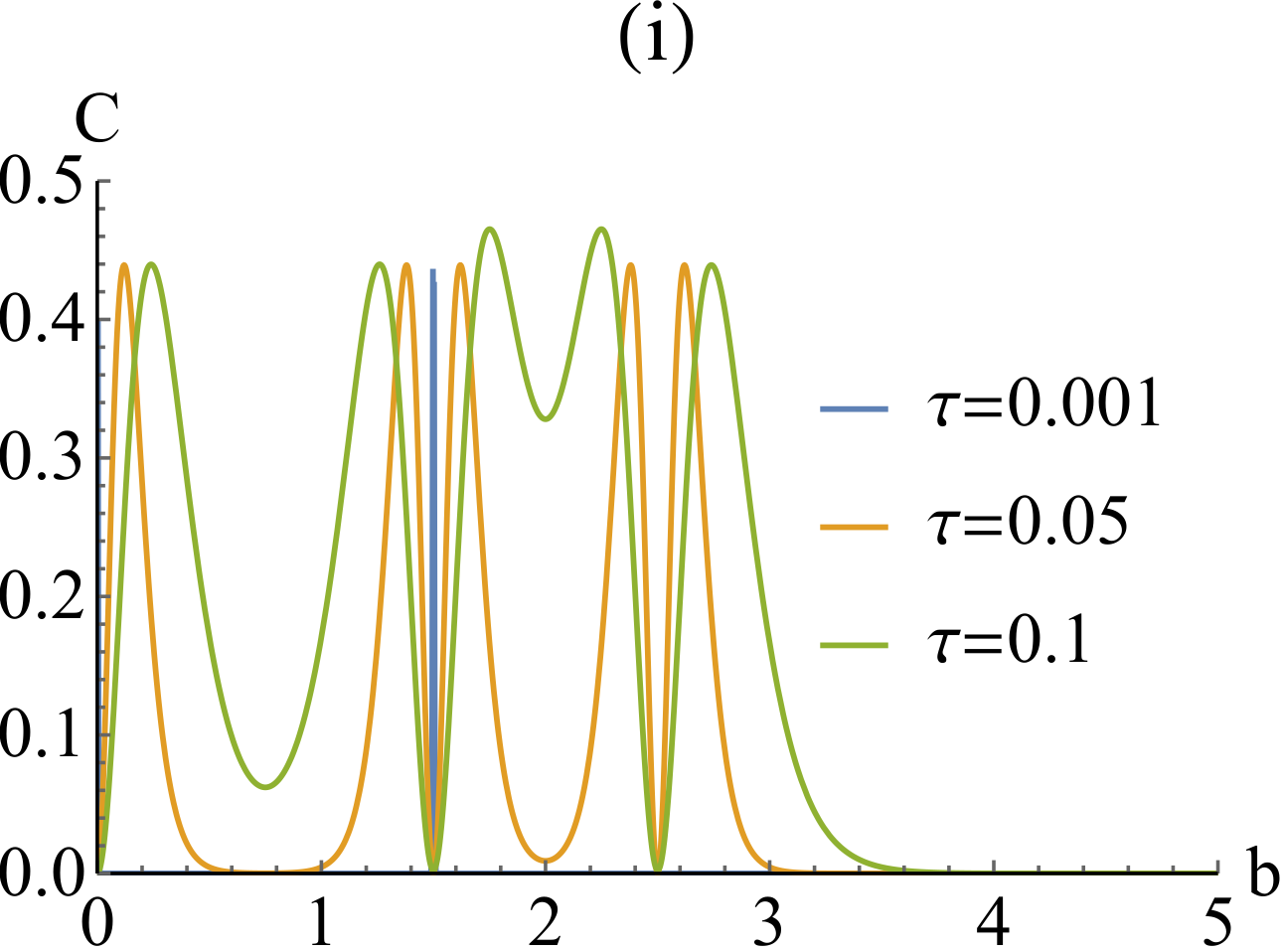}
\includegraphics[width=0.32\linewidth]{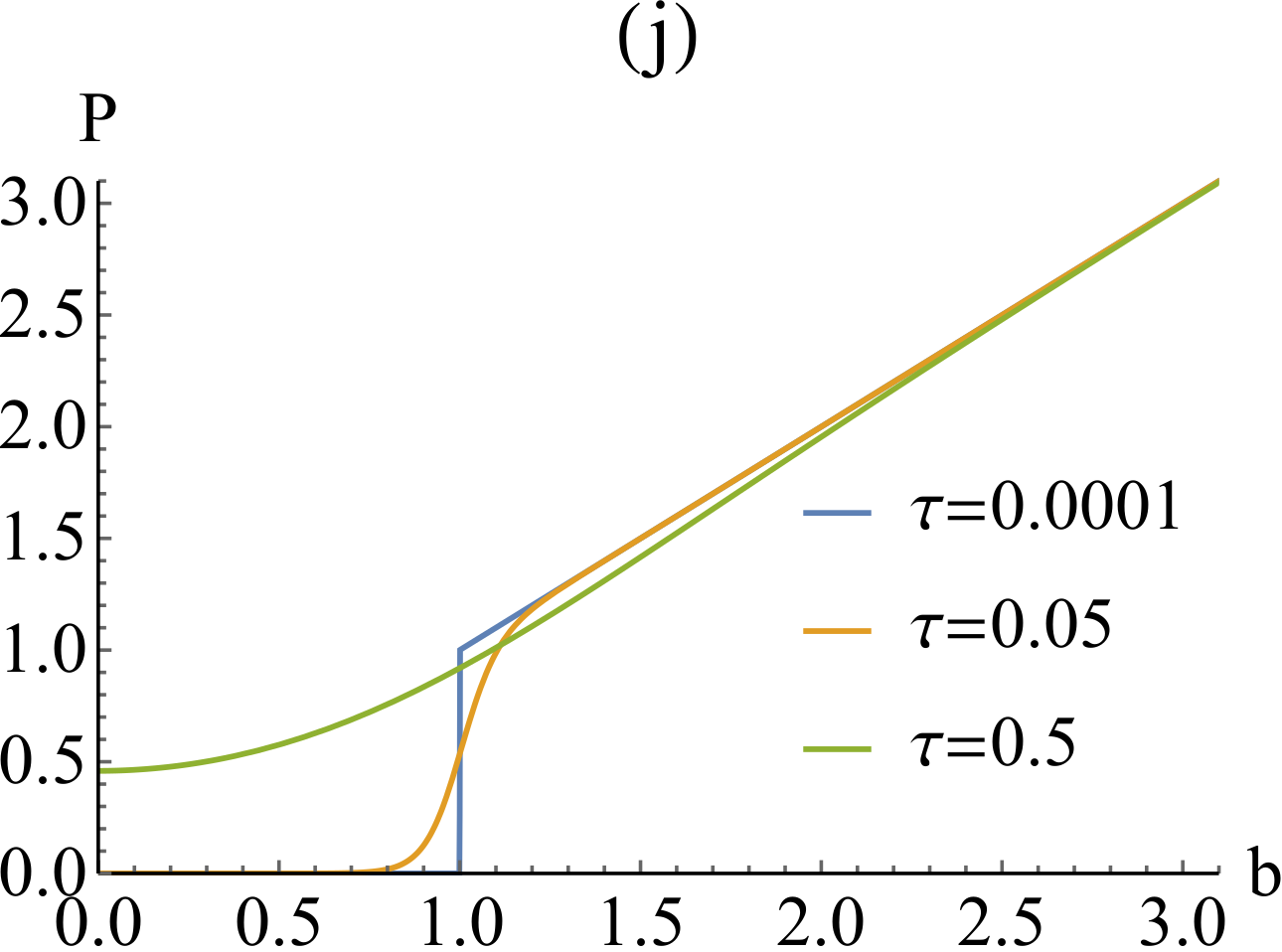}
\includegraphics[width=0.32\linewidth]{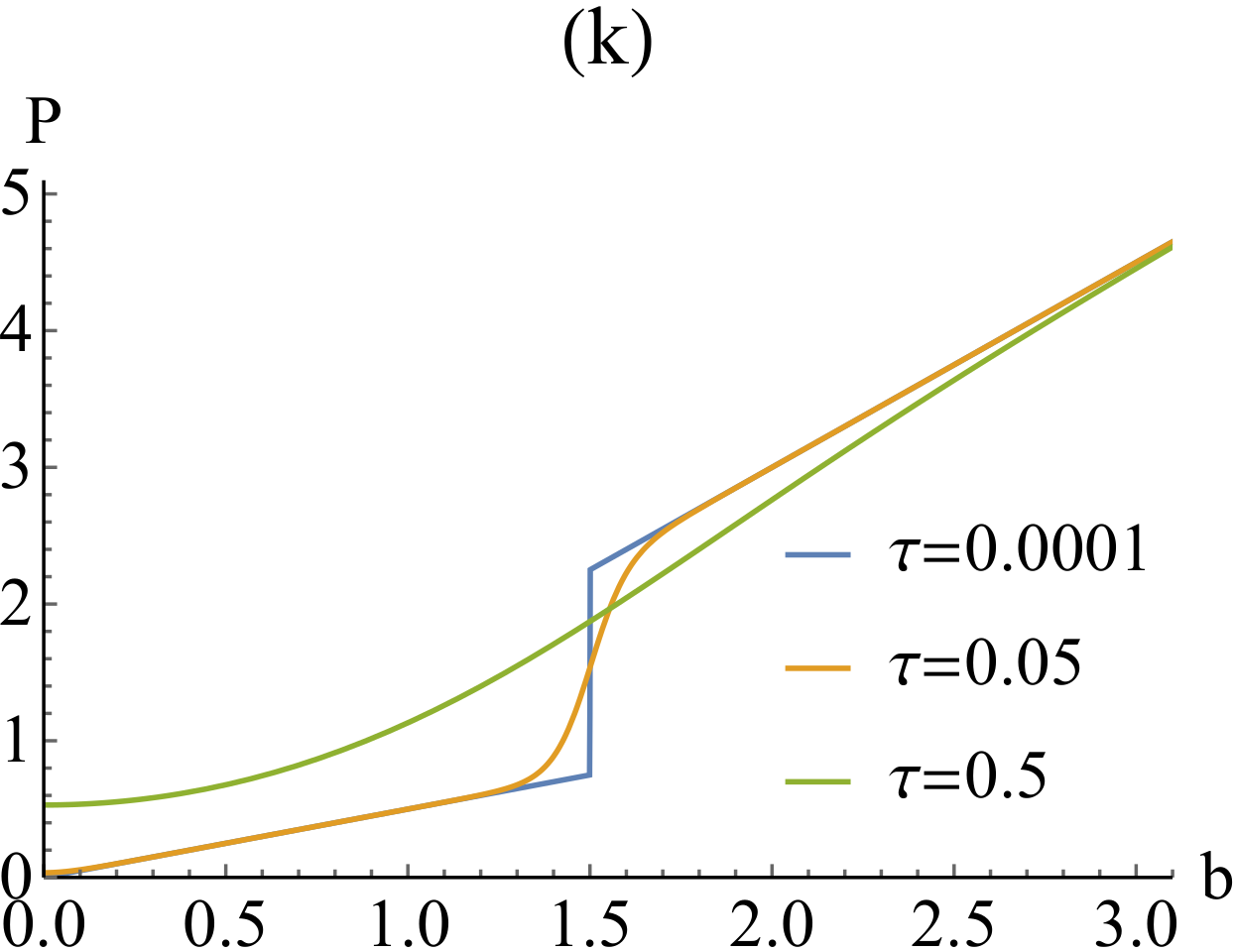}
\includegraphics[width=0.32\linewidth]{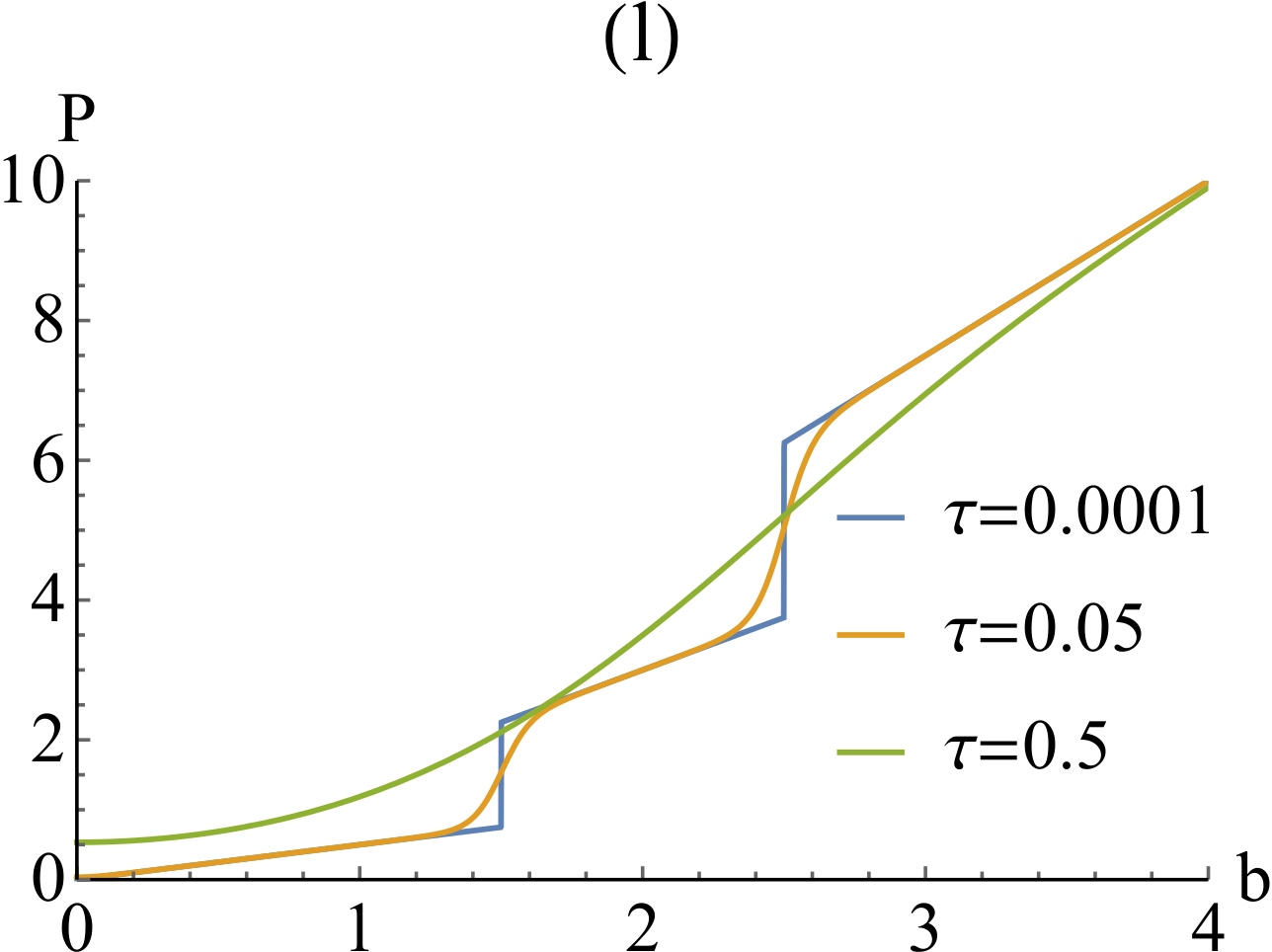}
\caption{\label{fig:SmallN}
Thermodynamic quantities for $N = 2,3,5$ particles.
(a),(d),(g),(j) are for $N = 2$, (b),(e),(h),(k) for $N=3$, and (c),(f),(i),(l)
for $N=5$.  The first row ((a)-(c)) shows the magnetization per particle $m$
vs.\ field $b$ for three temperatures $\tau = 0.0001,0.1,0.5$.  The second
row ((d)-(f)) shows the linear susceptibility $\chi_1$ and the first two
nonlinear susceptibilities $\chi_3,\chi_5$ vs.\ temperature $\tau$.  Both (e)
and (f) have subtracted out the free spin response contribution to the
susceptibilities.  The third row ((g)-(i)) shows the specific heat $C$ vs.\
field $b$ for three temperatures $\tau = 0.001,0.05,0.1$.  The final row
((j)-(l)) shows the pressure $P$ vs.\ field $b$ for three temperatures
$\tau = 0.0001, 0.05, 0.5$.
}
\end{figure*}

It is possible to rewrite the magnetization for odd particle number
$M(N = 2n+1)$ in order to highlight the free spin term.  This calculation
gives a term similar to what is seen for $M(N = 2n)$ in Eq.\ \eqref{eqn:M},
%In order to highlight the free spin term in the magnetization for odd particle
%number, it is possible to rewrite $M(N = 2n + 1)$ in terms of $M(N-1 = 2n)$.
%When doing this the field independent function $A_\mu(N-1)$ in $M(N-1 = 2n)$
%is replaced with a new function $B_{\mu +1/2}(N)$,
\begin{widetext}
\begin{equation}
%\begin{align}
\frac{M(N = 2n + 1)}{\gamma} = \frac{1}{2} \tanh\left( \frac{b}{2\tau} \right) + \frac{2 \sum_{\mu=3/2}^{N/2} (\mu-1/2) B_\mu(N) \sinh((\mu-1/2)b/\tau)}{B_{1/2}(N) + 2 \sum_{\mu=3/2}^{N/2} B_\mu(N) \cosh((\mu-1/2)b/\tau)},
%\frac{M(N = 2n + 1)}{\gamma} &= \frac{1}{2} \tanh\left( \frac{b}{2\tau} \right) + \frac{2 \sum_{\mu=3/2}^{N/2} (n-1/2) B_n(N) \sinh((n-1/2)b/\tau)}{B_{1/2}(N) + 2 \sum_{\mu=3/2}^{N/2} B_n(N) \cosh((n-1/2)b/\tau)}, \nonumber \\
%&= \frac{1}{2} \tanh\left( \frac{b}{2\tau} \right) + \frac{1}{\gamma} M(N-1=2n,A_\mu(N-1) \rightarrow B_{\mu + 1/2}(N)),
%\frac{M(N = 2n + 1)}{\gamma} &= \frac{1}{2} \tanh\left( \frac{b}{2\tau} \right) + \frac{1}{\gamma} M(N-1=2n,A_\mu(N-1) \rightarrow B_{\mu + 1/2}(N)),
\label{eqn:Modd}
%\end{align}
\end{equation}
\end{widetext}
where $B_\mu(N)$ is defined from the functions $A_m(N)$,
\begin{equation}
B_\mu(N) = 2 \sum_{m = \mu}^{N/2} (-1)^{m - \mu} A_m(N).
\label{eqn:Bparm}
\end{equation}
By reindexing the sums in Eq.\ \eqref{eqn:Modd} using $\mu' = \mu - 1/2$
and replacing $B_{\mu'+1/2}(N)$ with $A_{\mu'}(N-1)$ the second term
gives back the magnetization for even particle number.
The full derivation of the above equivalency can be seen in Appendix A.

\section{3. Small $N$ Limit}

At this stage it is important to look into expressions for specific values of
$N$, both to get a better, more comprehensive understanding and also to check
the validity of some results in the space of hyperbolic functions.  Some of
these results can be derived by direct calculation of the partition function
and used as a check.

\begin{center}
(a) $N = 2:$
\end{center}
Two spin half particles are described by a partition function
$Z = 1 + e^{-1/\tau} (1 + 2 \cosh b/\tau)$.  The magnetization $M$ is given by
\begin{equation}
M = \gamma \frac{\sinh(b/\tau)}{C_2 + \cosh(b/\tau)}, \quad \quad C_2 = \frac{1}{2} A_0(2) = \frac{1}{2} ( 1 + e^{1/\tau}).
\label{eqn:M(2)}
\end{equation}
which leads to
\begin{subequations}
\begin{align}
\chi_1(T) &= \frac{\gamma^2}{k T} \frac{1}{1 + C_2}, \label{eqn:chi1(2)} \\
\chi_3(T) &= \frac{\gamma}{3!} \left(\frac{\gamma}{k T} \right)^3 \frac{C_2 - 2}{(1 + C_2)^2}, \label{eqn:chi3(2)} \\
\chi_5(T) &= \frac{\gamma}{5!} \left( \frac{\gamma}{k T} \right)^5 \frac{C_2^2 - 13 C_2 + 16}{(1 + C_2)^3}. \label{eqn:chi5(2)}
\end{align}
\end{subequations}

\begin{figure}
\includegraphics[width=0.49\linewidth]{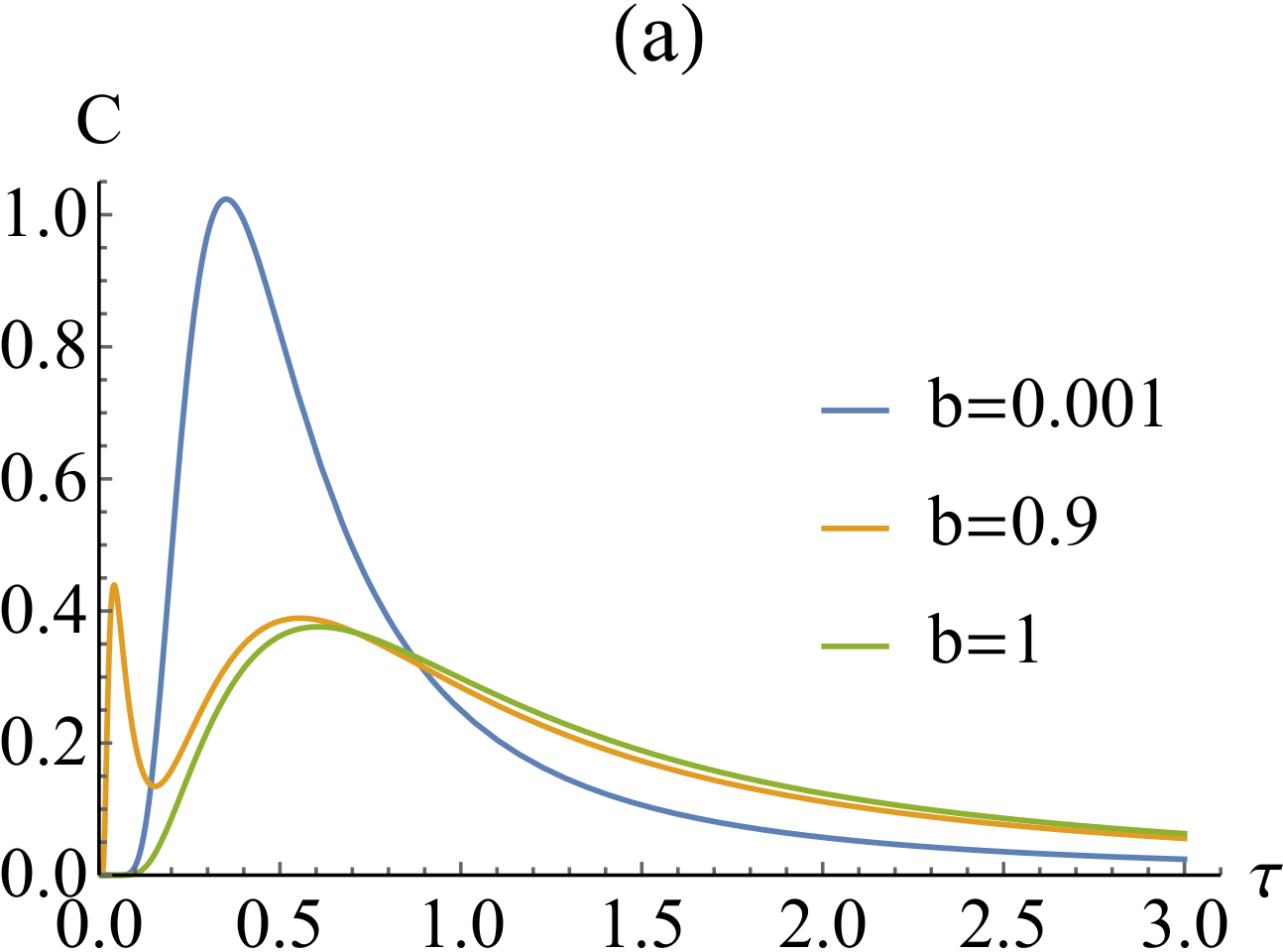}
\includegraphics[width=0.49\linewidth]{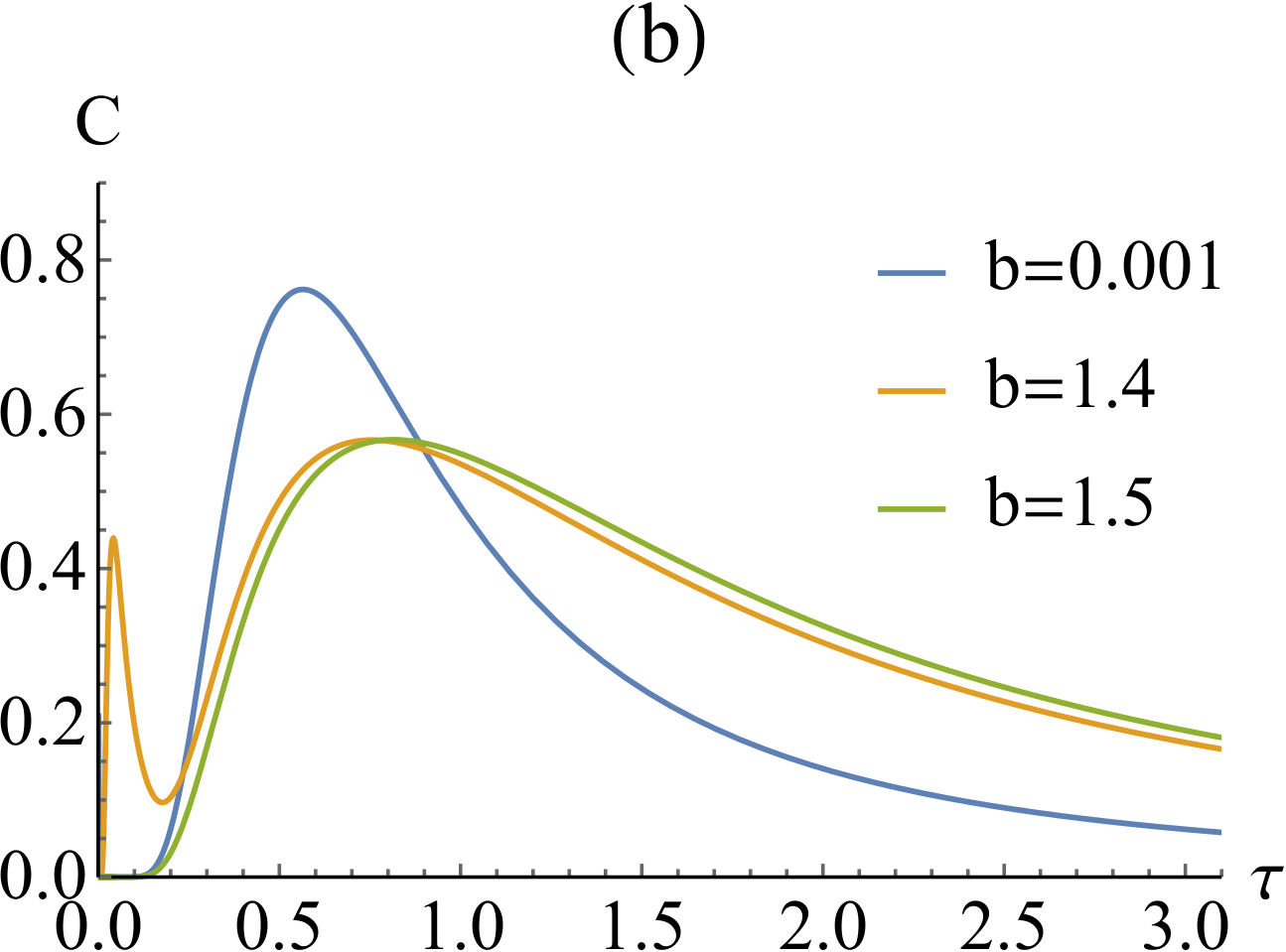}
\includegraphics[width=0.49\linewidth]{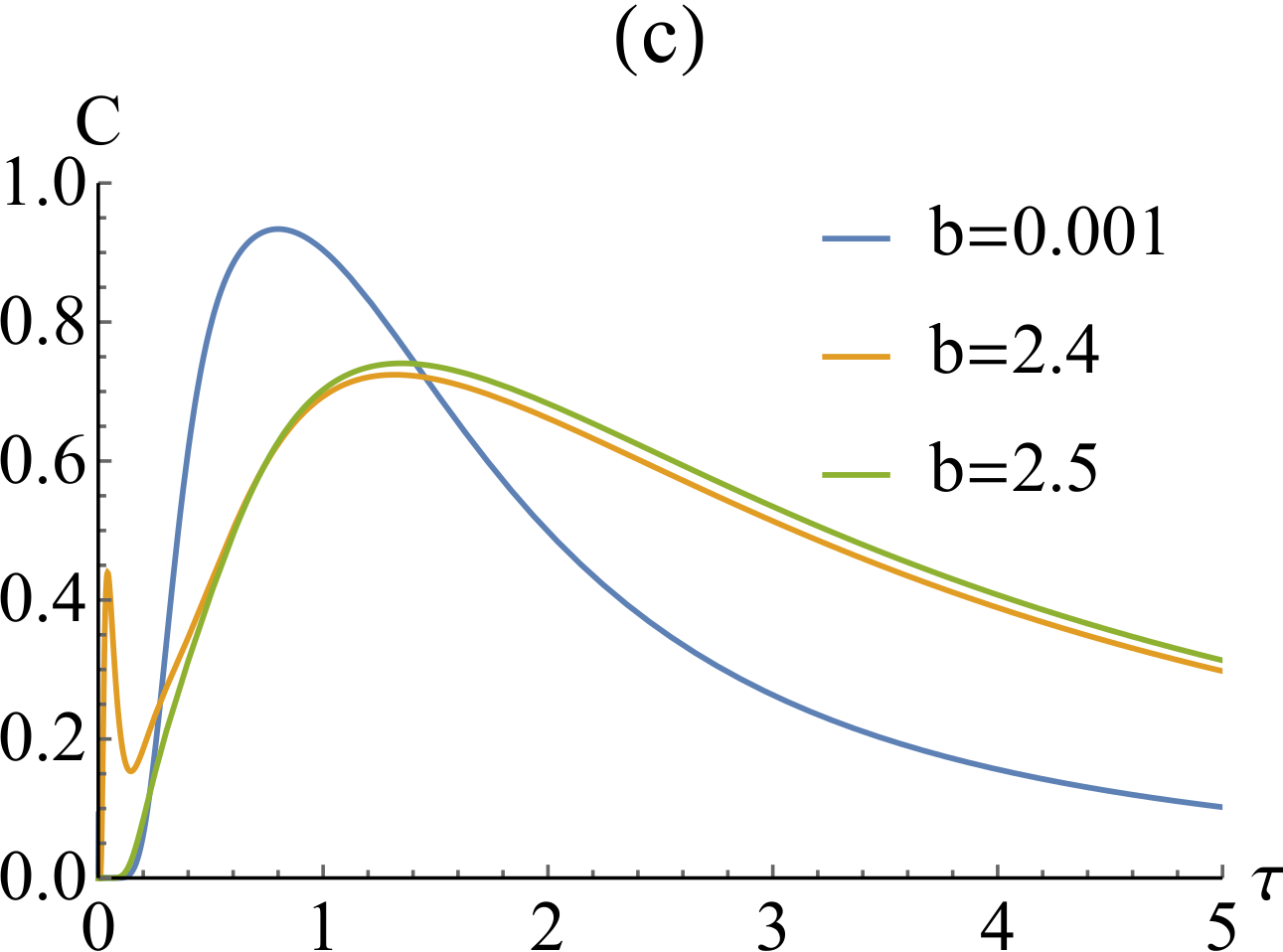}
\caption{\label{fig:CSmallN}
Specific heat versus temperature for $N = 2, 3, 5$.  (a) The specific heat
for $N = 2$ for field values $b = 0.001, 0.9, 1$.  These values show the
change from small $b$ to the humps that occur at small $\tau$ for
$b \sim b_c$.  (b) The specific heat for $N = 3$ for field values
$b = 0.001, 1.4, 1.5$.  Humps form similarly to the $N = 2$ case.  (c) The
specific heat for $N = 5$ for field values $b = 0.001, 2.4, 2.5$.  These
plots emphasize the humps occurring for small temperatures $\tau$ when
$b \sim b_c$.
}
\end{figure}

These are the principal nonlinear susceptibilities.  The results for the linear susceptibility \cite{Bleaney:52} and zero temperature magnetization
\cite{Tachiki:70} are well known.  The results are shown in Figs.\
\ref{fig:SmallN}(d).  The temperature dependence of the susceptibilities
is qualitatively similar to the anisotropy based models \cite{Shivaram:14}
(replacing $C_2 \rightarrow \frac{1}{2} e^{1/\tau}$).  The linear susceptibility
$\chi_1$ has a maximum at $T_1 = 0.624J$ and vanishes at both low and high
temperatures.  At high temperatures, it has an effective Curie-Weiss
temperature $\theta = J/4$.  The third order susceptibility $\chi_3(T)$
is negative at high temperatures but has a positive maximum at
$T_3 = 0.265J$ and vanishes at $T = 0$.  The next order nonlinear
susceptibility, $\chi_5(T)$ is qualitatively similar with a peak at
$T_5 = 0.176J$.

The specific heat for $N = 2$ is shown versus magnetic field in Fig.\
\ref{fig:SmallN}(g) and versus temperature in Fig.\ \ref{fig:CSmallN}(a).  As
a function of temperature, the specific heat has the usual Schottky features,
e.g.\ those of the specific heat of a two level system.  They include an
exponential rise at low $T$ and an inverse power law decay at high
temperatures.  As a function of magnetic field for low $T$ in Fig.\
\ref{fig:SmallN}(g), it has the $M$ shaped response centered at $b = 1$.  As
the temperature increases, the specific heat at $b = 0$ increases, the two
peaks move away from each other, but the minimum stays around $b = 1$.

The magnetic field dependent part of the pressure is shown in Fig.\
\ref{fig:SmallN}(j).  It consists of a threshold proportional to temperature
at $b = 1$ and a linear dependence on $b$ for $b > 1$.

\begin{center}
(b) $N = 3,5:$
\end{center}
Here the ground state is a Kramer's doublet.  This leads to several interesting
effects.  The partition function for $N = 3$ is given by:
\begin{equation}
Z = 2 e^{-3/8\tau} \cosh\left( \frac{b}{2\tau} \right) \left[ 1 + 2 e^{3/2\tau} \cosh\left( \frac{b}{\tau} \right) \right]
\label{eqn:Z(3)}
\end{equation}
and the magnetization $M$ is given by
\begin{align}
\begin{split}
M &= \frac{\gamma}{2} \tanh \left( \frac{b}{2\tau} \right) + \gamma \frac{\sinh(b/\tau)}{C_3 + \cosh(b/\tau)}, \\
C_3 &= \frac{1}{2} B_{1/2}(3) = \frac{1}{2} e^{3/2\tau}.
\end{split}
\label{eqn:M(5)}
\end{align}
The first term is the free particle $S = 1/2$ response.  With the odd number
of particles this is the dominant contribution at low temperatures.  The
linear and nonlinear susceptibilities are given by
\begin{subequations}
\begin{align}
\chi_1 &= \frac{\gamma^2}{4 k T} \left[ 1 + \frac{4}{1 + C_3} \right], \label{eqn:chi1(3)} \\
\chi_3 &= \frac{\gamma}{3} \left( \frac{\gamma}{2 k T} \right)^3 \left[ 1 + 2^3 \frac{C_3 - 2}{(1 + C_3)^2} \right], \label{eqn:chi3(3)} \\
\chi_5 &= \frac{2 \gamma}{15} \left( \frac{\gamma}{2 k T} \right)^5 \left[ 1 + 2^5 \frac{5 C_3^2 - 13 C_3 + 16}{(1 + C_3)^3} \right]. \label{eqn:chi5(3)}
\end{align}
\end{subequations}
The first term in each of the above equations is the free particle $s = 1/2$
response.  The second term is (and the following terms are) the usual linear
and nonlinear susceptibility albeit with an $N$ dependent $C_N$.  There is a
step at $b = 0$ leading to a magnetization per particle $m = 1/6$.  This is
followed by a step at $b = 3/2$ with $m = 1/2$.  For odd number of particles
without a magnetic field the ground state is doubly degenerate and the
magnetization vanishes.  However at the smallest field there is a separation
in the energy levels for the $m = \pm 1/2$ leading to a nonzero magnetization
at low $T$ and a step in $m(b)$ at $b = 0$.

The general features for an odd $N$ are seen in Figs.\ \ref{fig:SmallN} for
both $N = 3$ and $N=5$.  The $N=5$ plots more clearly show these features.
Figs.\ \ref{fig:SmallN}(b),(c) show the magnetization steps at the critical
values of the magnetic field.  The step at $b = 0$ is followed by one at
$b = 3/2$ (for $N=3,5$) and at 5/2 (for $N=5$).  Figs.\
\ref{fig:SmallN}(e),(f) show the linear and nonlinear susceptibilities with
the low temperature limit with the free spin response contribution removed.
When included, all $\chi_i$ are divergent for the $\tau \rightarrow 0$ limit.
The features are similar to the even $N$ response characteristic of the model;
negative at high $T$.  As in the even $N$ case the specific heat (Figs.\
\ref{fig:SmallN}(h),(i)) dips to zero at the critical field values for small
temperatures.  The temperature dependence of the specific heat can be seen in
Figs.\ \ref{fig:CSmallN}(b),(c) for $N=3,5$.  The pressure as a function of
magnetic field (Figs.\ \ref{fig:SmallN}(k),(l)) show the phase transition at
the critical field values, along with a slope equal to the ground state spin
for small $\tau$.

\section{4. Large $N$ Limit}

The partition function in Eq.\ \eqref{eqn:Z} has a large $N$ limit which can
be studied in two alternative ways.  Analytically, an infinite $N$ limit can
be studied by replacing the sum by an integral.  The magnetization turns into
a Gaussian integral of the form:
\[
M = \int_{n_0/2}^\infty dS e^{-S(S+1)/2} f(S), \quad f(S) = \frac{\sinh\left(\frac{(2S + 1)b}{2\tau} \right)}{\sinh(b/2\tau)},
\]
which can be evaluated in terms of error functions.  The results can be made
more transparent by evaluating the sums directly over a large number of
particles, such as $N = 23$ or 24 (odd and even cases separately), and
interpolating $N \rightarrow \infty$.

\begin{figure*}
\includegraphics[width=0.32\linewidth]{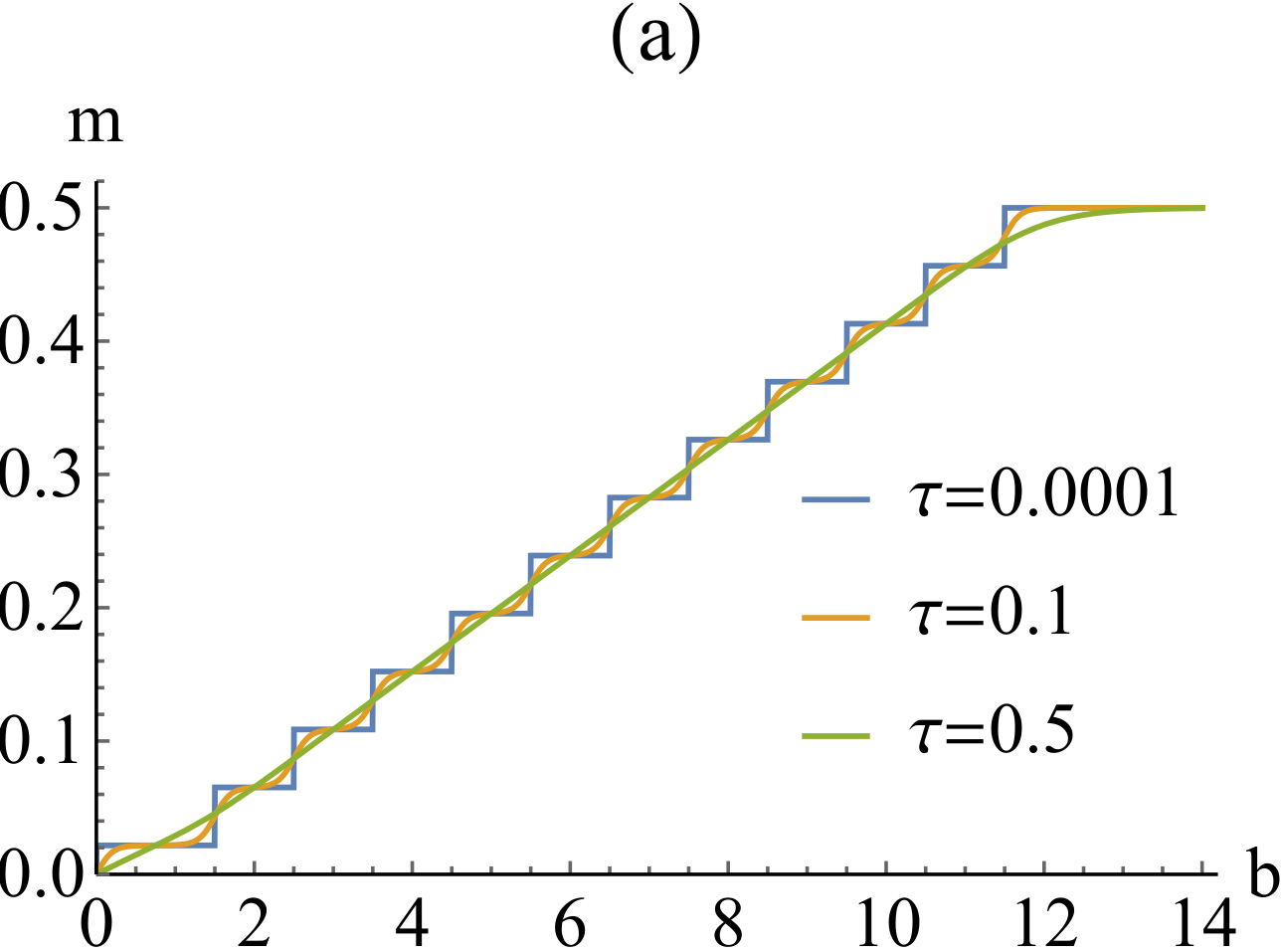}
\includegraphics[width=0.32\linewidth]{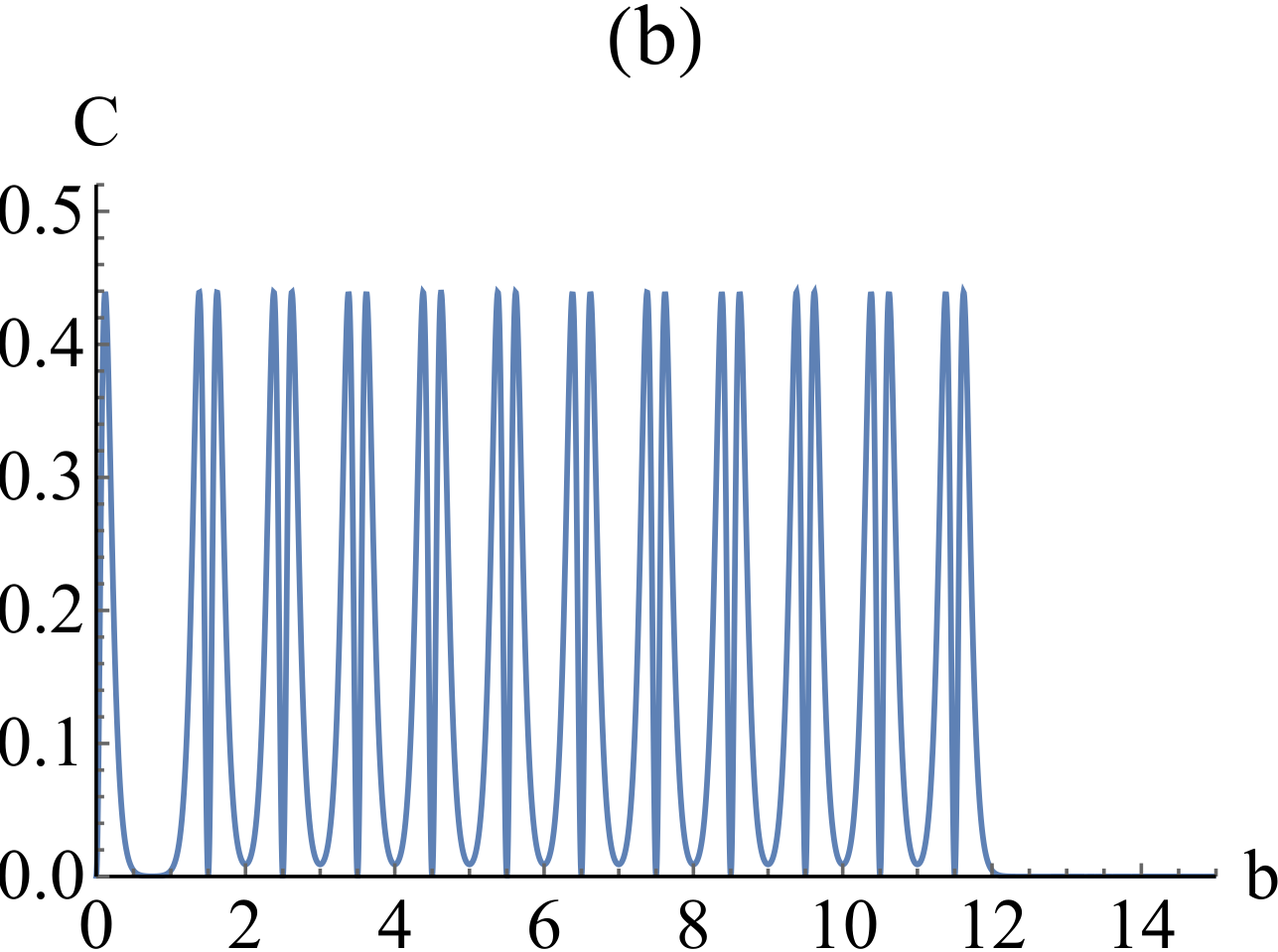}
\includegraphics[width=0.32\linewidth]{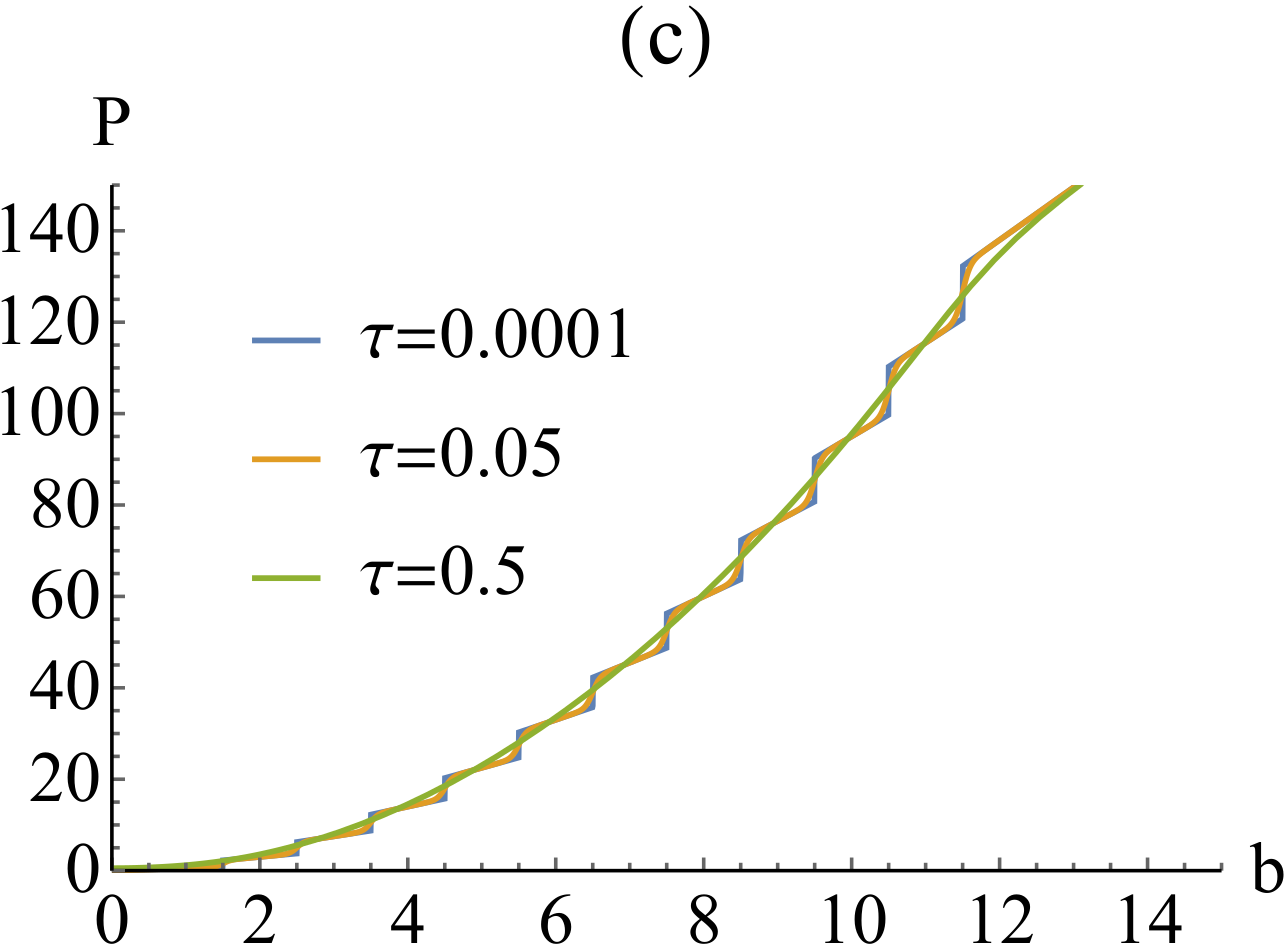}
\includegraphics[width=0.32\linewidth]{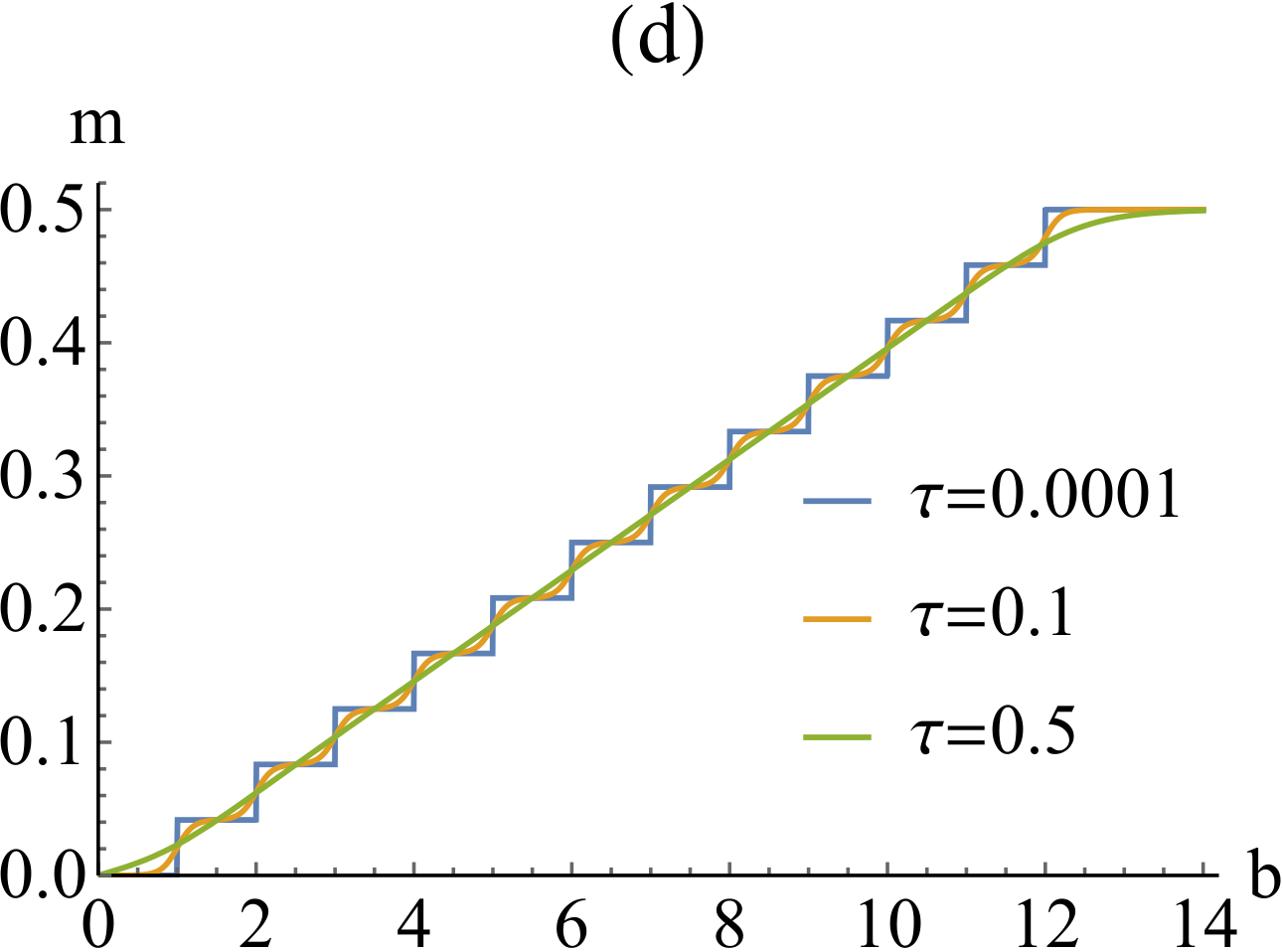}
\includegraphics[width=0.32\linewidth]{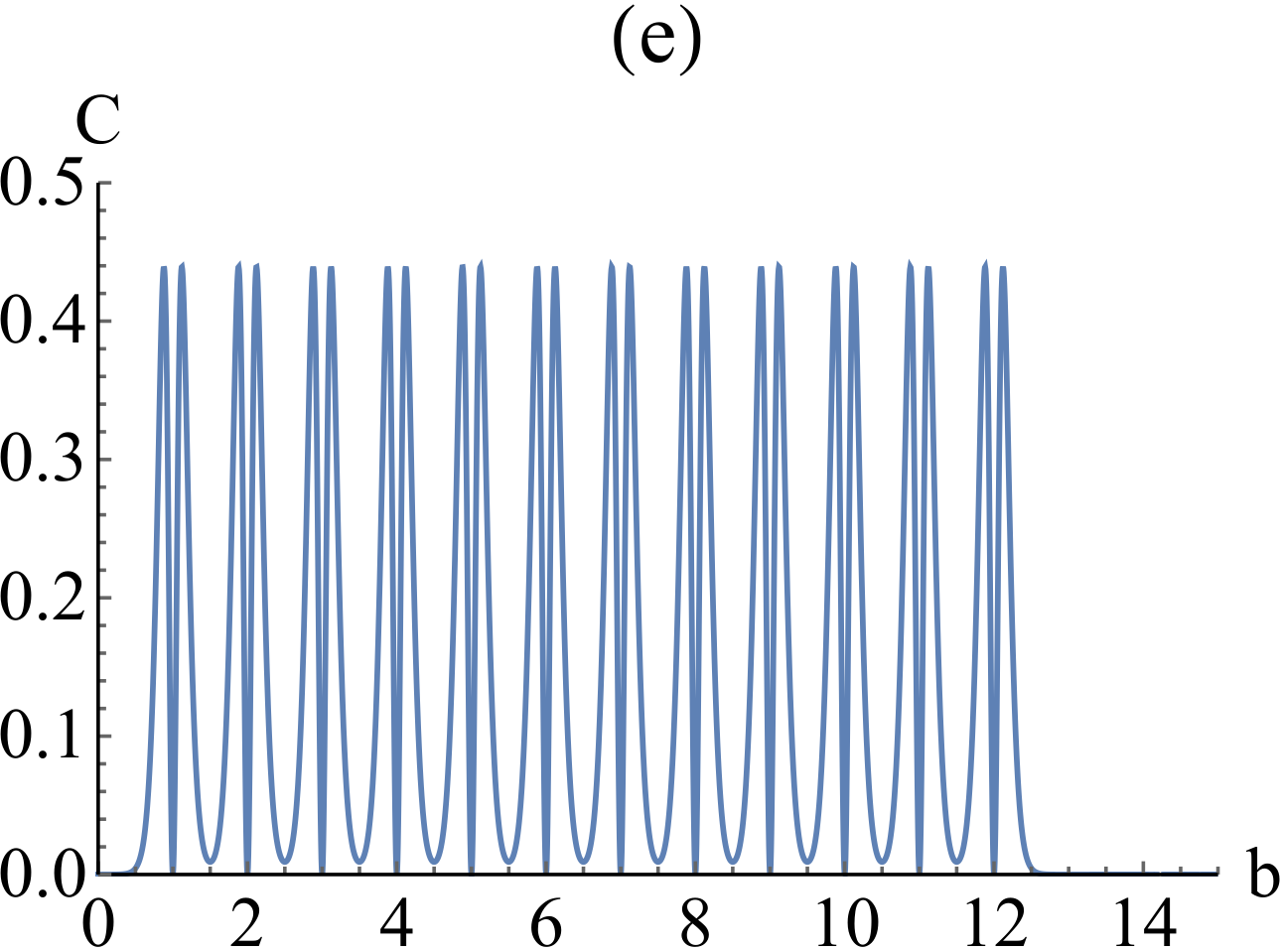}
\includegraphics[width=0.32\linewidth]{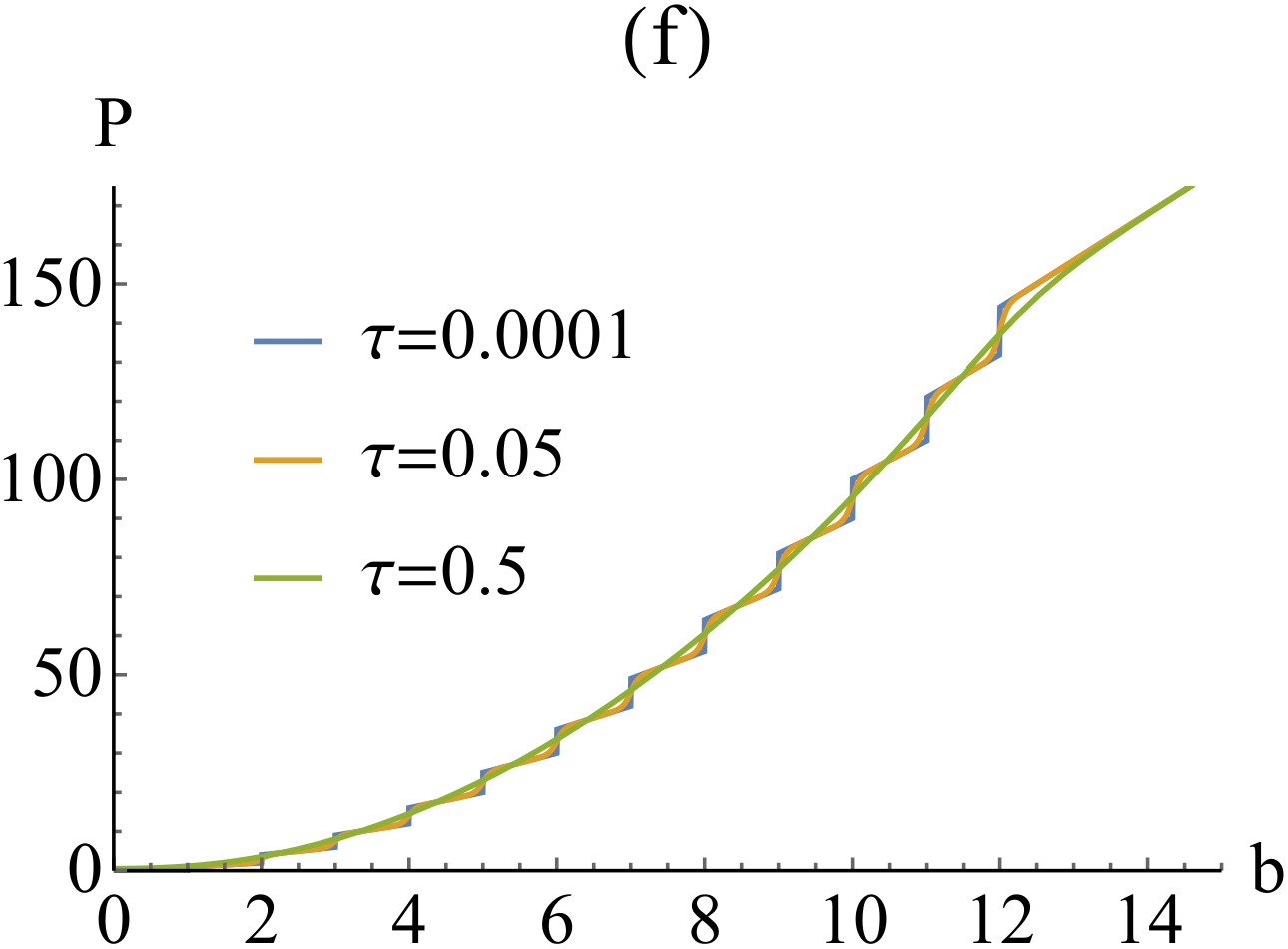}
\caption{\label{fig:LargeN}
Thermodynamic quantities for $N=23,24$.  (a)-(c) are for $N=23$ and $(d)-(f)$
are for $N=24$.  The first column ((a),(d)) are the magnetization per particle
$m$ vs.\ field $b$ for three temperatures $\tau = 0.0001,0.1,0.5$.  The second
column ((b),(e)) shows the M shape in the specific heat $C$ vs.\ field $b$ for
$\tau = 0.05$.  The third column ((c),(f)) shows the pressure $P$ vs.\ field
$b$ for three temperatures $\tau = 0.0001, 0.05, 0.5$.
}
\end{figure*}

Fig.\ \ref{fig:LargeN} shows the full extension of the model to large particle
number.  Fig.\ \ref{fig:LargeN}(a) plots the magnetization per particle $m$ as
a function of magnetic field at low temperatures for $N = 23$.  There is the
expected step at $b = 0$ followed by 11 more steps at
$b = 3/2, 5/2, ..., 23/2$, which can be seen in both the magnetization per
particle and the pressure in Fig.\ \ref{fig:LargeN}(c).  The M shape in the
specific heat (Fig.\ \ref{fig:LargeN}(b)) at small temperatures can be seen
for each of the critical field values.  The even $N$ case for the
magnetization per particle, specific heat, and pressure using $N = 24$ are
seen in Figs.\ \ref{fig:LargeN}(d)-(f).  Features similar to the $N = 23$ case
are seen for $N = 24$.

The results here show that there will be steps at integer $b$ in the
magnetization per particle and pressure which quickly blur with increased
temperature.  This implies an infinite number of steps in the thermodynamic
limit.  Since the change in the magnetization is the same in all cases, this
means that the change in the magnetization per particle is always $1/N$,
except for the initial $b=0$ change for odd $N$, which is $1/2N$.  In the
thermodynamic limit these steps are infinitesimally small for both even and
odd $N$, implying that in the thermodynamic limit this model shows no
metamagnetism.

\section{5. Discussion and Conclusions}

Metamagnetism in a strongly correlated metal derives its properties from the
complicated interactions.  It is usually described by a lattice of magnetic
moments interacting with conduction electrons and often within a framework of
Anderson model.  The heavy mass of fermions is then an outcome of a Kondo-like
effect.  The ground state corresponds to a singlet resonance between the local
moment and the conduction electrons.  Metamagnetism happens when the magnetic
field breaks the singlet resonance.  This scenario is the basis of the
microscopic calculations described \cite{Brenig:87, Ono:98} in several
references.  Several experimental correlations \cite{Aoki:13, Hirose:11}
appear reasonable within this formalism.  One outcome that remains to be
explored experimentally is that between the critical field and the effective
mass.

However in order to calculate a more complex observable, such as nonlinear
susceptibilities, a truly microscopic calculation is a non-starter.  An
intermediate framework needs to be established which both transcends
microscopic parameters and has the facility to proceed with more complex
calculations.  Described here is an orthogonal approach, studying a model that
seems reasonable from a microscopic point of view but is simple enough to
yield nonlinear susceptibilities.

The infinite range interaction model contains many of the properties one finds
in metamagnetism.  The model contains only one energy scale, $J$, apart from
the thermodynamic control variables, the external magnetic field and the
temperature. It is solvable with transparent intermediate steps.
The notable features of this model are:
\begin{enumerate}[(a)]

\item There is a quantum transition at $T = 0$ that loses its singular
transition properties at any non-zero temperature.  Thus the $T = 0$ transition
is discontinuous in magnetization as well as in entropy.  The total magnetic
susceptibility $\chi(B, T=0)$ is singular at $B = B_c$.  At any finite
(non-zero) temperature, the magnetic susceptibility has maxima at $B = B_c$.

\item The nonlinear susceptibilities defined by
\[
M = \chi_1(T) B + \chi_3(T) B^3 + \chi_5(T) B^5.
\]
are negative at high temperatures.  They be come positive at low temperatures
and go through a maximum.

\item There are features due to the oddness of the total number of particles
$N$.  For odd $N$, there is a free spin contribution to the susceptibility
that diverges as $1/T$.

\item Finally there are results for the field dependence of the specific heat
and pressure (from which the magnetic field dependence of the sound velocity
can be derived).

\end{enumerate}

The model is, in the classical sense, highly frustrated \cite{Chalker:11}.
In the thermodynamic limit, there are no phase transitions as a function of
temperature.  This is due to the infinitesimal step size from an infinite
number of transitions in $M$ in the thermodynamic limit.  The model is a
spin model that sidesteps the complexities of a real microscopic model for
a strongly correlated fermion system.  For a ferromagnetic version of this model ($J < 0$) see Ref.\ \footnote{Notes on the ferromagnetic ($J < 0$) version of this model can be viewed in J.\ J.\ Binney \textit{et al}, \textit{Theory of Critical Phenomena}, Clarendon Press, NY (1993) and in notes by M.\ C.\ Cross online at www.cmp.caltech.edu/$\sim$mcc/BNU/Notes2\_4.pdf}.  The objective here has been to
develop a minimal (spins only) model which highlights the common features
among materials with different lattice structures and electronic properties.
The model may be useful in the fields of quantum magnetism \cite{Alcaraz:88} and molecular magnets \cite{Schnack:13}.

We acknowledge several stimulating and enlightening discussions with V.\ Celli,
B.\ Cowan, B.\ Nartowt and B.\ Shivaram.

%Bibliography

\appendix
\begin{widetext}
\section{Appendix A: Magnetization for odd $N$}
The free spin term of the magnetization for odd particle number is able to
be extracted (as in Eq.\ \eqref{eqn:Modd}) from $M(N=2n+1)$ in
Eq.\ \eqref{eqn:M}.  The derivation of this is contingent upon three steps:
\begin{enumerate}[1.)]

\item \underline{Rewrite the partition function
$Z(N=2n+1) = \cosh(b/2\tau) F(b/\tau)$.}

The definition of $Z(N=2n+1)$ given in Eq.\ \eqref{eqn:modZ} is:
\[
Z(N=2n+1) = 2 \sum_{\mu = 1/2}^{N/2} A_\mu(N) \cosh \left( \frac{\mu b}{\tau} \right).
\]
Since $A_\mu(N)$ is field independent, then the $\cosh(b/2\tau)$ must
come from the $\cosh(\mu b/\tau)$ terms.  Using the hyperbolic angle sum rules
for $\cosh(x + y)$ and $\sinh(2x)$:
\[
\begin{split}
\cosh(nb/\tau) &= \cosh((n-1)b/\tau) \cosh(b/\tau) + \sinh((n-1)b/\tau) \sinh(b/\tau), \\
&= \cosh((n-1)b/\tau) \cosh(b/\tau) + 2 \cosh(b/2\tau) \sinh((n-1)b/\tau) \sinh(b/2 \tau).
\end{split}
\]
Assuming that $\cosh((n-1)b/\tau) = \cosh(b/2\tau) f_{n-1}(b/\tau)$,
then the above becomes:
\[
\begin{split}
\cosh(nb/\tau) &= \cosh((n-1)b/\tau) \cosh(b/\tau) + 2 \cosh(b/2\tau) \sinh((n-1)b/\tau) \sinh(b/2 \tau), \\
&= \cosh(b/2\tau) [ f_{n-1}(b/\tau) \cosh(b/\tau) + 2 \sinh((n-1)b/\tau) \sinh(b/2 \tau) ].
\end{split}
\]
The square brackets must then be $f_n(b/\tau)$, giving a recursion relation
for $f_n(b/\tau)$:
\begin{equation}
f_n(b/\tau) = f_{n-1}(b/\tau) \cosh(b/\tau) + 2 \sinh((n-1)b/\tau) \sinh(b/2\tau).
\label{eqn:fn}
\end{equation}
By showing that this process is started using $n = 1/2$ and $n = 3/2$, the
recursion can be proven true through induction.  Using the relation
$\cosh(nb/\tau) = \cosh(b/2\tau) f_n(b/\tau)$, the partition function can
be rewritten:
\begin{equation}
Z(N=2n+1) = \cosh(b/2\tau) \left[ 2\sum_{\mu=1/2}^{N/2} A_\mu(N) f_\mu(b/\tau) \right] = \cosh(b/2\tau) F(b/\tau).
\label{eqn:ZF}
\end{equation}

\item \underline{Find a closed form for $F(b/\tau)$.}

The only unknown in $F(b/\tau)$ from Eq.\ \eqref{eqn:ZF} is the closed form
of $f_\mu(b/\tau)$.  This closed form is not obvious from Eq.\ \eqref{eqn:fn}.
By rearranging the terms for $f_{3/2}(b/\tau)$ and $f_{5/2}(b/\tau)$, they
can be rewritten using a nicer recursion relation:
\begin{equation}
f_n(b/\tau) = 2 \cosh((n-1/2)b/\tau) - f_{n-1}(b/\tau),
\label{eqn:fnew}
\end{equation}
which has a straightforward closed form solution,
\begin{equation}
f_n(b/\tau) = 2 \sum_{m = 0}^{n-1/2} (-1)^m \cosh((n-1/2-m)b/\tau) - (-1)^{n-1/2}
\label{eqn:fnclosed}
\end{equation}
If this is the correct closed form for $f_n(b/\tau)$, then it will obey the
original recursion of Eq.\ \eqref{eqn:fn}.  Plugging
Eq.\ \eqref{eqn:fnclosed} into the original recursion Eq.\ \eqref{eqn:fn}
and using the relation $\cosh(x)\cosh(y) = \cosh(x+y) - \sinh(x)\sinh(y)$
gives:
\[
\begin{split}
f_n(b/\tau) &= \cosh(b/\tau) f_{n-1}(b/\tau) + 2 \sinh(b/2\tau)\sinh((n-1)b/\tau), \\
&= 2 \cosh((n-1/2)b/\tau) - \underbrace{2 \sum_{m=1}^{n-3/2} (-1)^{m-1} \cosh((n-3/2-(m-1))b/\tau)}_{=f_{n-1}(b/\tau) - (-1)^{n-3/2}} + (-1)^{n-1/2} \cosh(b/\tau) \\
& \quad - 2\sum_{m=0}^{n-3/2} (-1)^m \sinh((n-3/2-m)b/\tau) \sinh(b/\tau) + 2 \sinh(b/2\tau)\sinh((n-1)b/\tau), \\
&= 2 \cosh((n-1/2)b/\tau) - f_{n-1}(b/\tau) + S_n,
\end{split}
\]
where the extra term $S_n$ is
\[
S_n = (-1)^{n-3/2} + (-1)^{n-1/2} \cosh(b/\tau) - 2\sum_{m=0}^{n-3/2} (-1)^m \sinh((n-3/2-m)b/\tau) \sinh(b/\tau) + 2 \sinh(b/2\tau)\sinh((n-1)b/\tau).
\]
This extra term is absent from Eq.\ \eqref{eqn:fnew}, implying that
$S_{n-1} = 0$.  It is straightforward to show that $S_n = 0$ by
considering $S_n = S_n + S_{n-1}$ and reducing.  Since $S_n = 0$, then
Eq.\ \eqref{eqn:fn} is the same recursion as Eq.\ \eqref{eqn:fnew}, meaning
that Eq.\ \eqref{eqn:fnclosed} is the closed form solution for $f_n(b/\tau)$.

\item \underline{Obtain the magnetization $M(N=2n+1)$}

Using Eq.\ \eqref{eqn:fnclosed} the partition function can be rewritten:
\begin{equation}
\begin{split}
Z(N=2n+1) &= \cosh(b/2\tau)\left[2\sum_{\mu=1/2}^{N/2} A_\mu(N) f_\mu(b/\tau) \right], \\
&= \cosh(b/2\tau)\left[ B_{1/2}(N) + 2 \sum_{\mu'=1}^{(N-1)/2} B_{\mu'+1/2}(N) \cosh(\mu' b/\tau) \right]. \\
\end{split}
\label{eqn:Zodd}
\end{equation}
The coefficients $B_n(N)$ are related to the coefficients $A_n(N)$:
\[
B_n(N) = 2 \sum_{m=n}^{N/2} (-1)^m A_m(N) \quad \longleftrightarrow \quad A_n(N) = \frac{1}{2} (B_n(N) + B_{n+1}(N)).
\]
This leaves the derivation of the magnetization $M(N=2n+1)$. This can be done
by taking the derivative of the natural log of the new form for the partition
function from above.  Taking this derivative:
%\[
%\begin{split}
\begin{equation}
\frac{M(N=2n+1)}{\gamma} = \tau \frac{\partial}{\partial b} \ln(Z(N)) = \frac{1}{2} \tanh(b/2\tau) + \frac{2 \sum_{\mu = 1}^{(N-1)/2} \mu B_{\mu+1/2}(N) \sinh(\mu b/\tau)}{B_{1/2}(N) + 2 \sum_{\mu=1}^{(N-1)/2} B_{\mu+1/2}(N) \cosh(\mu b/\tau)}.
%\frac{M(N=2n+1)}{\gamma} &= \tau \frac{\partial}{\partial b} \ln(Z(N)), \\
%&= \frac{1}{2} \tanh(b/2\tau) + \frac{2 \sum_{\mu = 1}^{(N-1)/2} \mu B_{\mu+1/2}(N) \sinh(\mu b/\tau)}{B_{1/2}(N) + 2 \sum_{\mu=1}^{(N-1)/2} B_{\mu+1/2}(N) \cosh(\mu b/\tau)}.
\end{equation}
%\end{split}
%\]
which is exactly the form provided in Eq.\ \eqref{eqn:Modd}.  The second
term resembles $M(N=2n)$ in Eq.\ \eqref{eqn:M} if $A_\mu(N=2n)$ is replaced
with $B_{\mu+1/2}(N=2n+1)$.

\end{enumerate}
\end{widetext}

\section{Appendix B: Steps in the magnetization}
For all cases presented, steps in the magnetization occur at integer values of
the dimensionless field $b = \gamma B/J$ for even particle number and
half-integer values of $b$ for odd particle number.  A change in the ground
state energy creates a change in the magnetization.  This is because the ground
state energy provides the largest weight to the partition function in the low
temperature limit $T \rightarrow 0$.  The critical field values can be
determined by when there is change in the ground state energy.

For an even number of particles, the eigenenergies are:
\[
\lambda = \frac{J}{2} S(S+1) - \gamma \mu B, \quad \quad 0 \le S \le N/2, \text{  }-S \le \mu \le S,
\]
where $S$ is the total spin of the system and $\mu$ is the $z$-component of
the total spin.  The constant term $3JN/8$ does not affect this proof and
thus was dropped.  For any given value of $S$, the smallest eigenenergy takes
$\mu = S$.  For $B = 0$ the smallest eigenenergy is simply $\lambda = 0$.  A
new ground state occurs when a higher energy level crosses this energy level.
The crossover locations for $S > 0$ with the $S = 0$ level are:
\[
0 = \frac{J}{2} S ( S + 1) - \gamma S B \quad \rightarrow \quad B = \frac{J}{2\gamma} (S + 1),
\]
The first crossing will create a new ground state energy level, occurring for
$S = 1$, or $B_c = \frac{J}{\gamma}$.  Another new ground state will occur
when a higher energy level ($S > 1$) crosses the $S = 1$ ground state energy
level.  All crossings can be calculated:
\[
\begin{split}
&\frac{J}{2} (1)(2) - \gamma (1) B = \frac{J}{2} S (S + 1) - \gamma S B. \\
\rightarrow &B = \frac{J}{2 \gamma} \frac{S (S + 1) - 2}{S - 1} = \frac{J}{2\gamma} (S + 2).
\end{split}
\]
The next highest spin ($S = 2$) will create the next ground state energy at
the crossing $B_c = 2 J/\gamma$.  When a ground state takes spin $S_g$ the
next ground state will have $S > S_g$ through induction:
\[
\begin{split}
&\frac{J}{2} (S_g)(S_g+1) - \gamma (S_g) B = \frac{J}{2} S (S + 1) - \gamma S B, \\
\rightarrow &B = \frac{J}{2 \gamma} \left( S + S_g + 1 \right).
\end{split}
\]
The next ground state will then always be for $S = S_g + 1$ at the crossing
$B_c = (S_g + 1) J/\gamma = n J/\gamma$.  Although this proof is for an even
particle number $N$, an equivalent proof can be performed for odd $N$ to show
the half integer crossings.  The value of the magnetization for
$T \rightarrow 0$ will be equal to the value of $S_g$ since
$M = \frac{1}{\beta} \partial_B \ln Z$ for small $T$ is approximated by the
ground state:
\[
\begin{split}
M(B,T \rightarrow 0) &\approx  \frac{1}{\beta} \frac{\partial_B e^{-\beta J S_g(S_g+1)/2 + \gamma \beta B S_g}}{e^{-\beta J S_g(S_g+1)/2 + \gamma \beta B S_g}}, \\
&\approx \gamma S_g \frac{e^{-\beta J S_g(S_g+1)/2 + \gamma \beta B S_g}}{e^{-\beta J S_g(S_g+1)/2 + \gamma \beta B S_g}}, \\
&\approx \gamma S_g.
\end{split}
\]

\end{document}